\documentclass[a4paper]{article}

\usepackage[english]{babel}
\usepackage[utf8x]{inputenc}
\usepackage[T1]{fontenc}

\usepackage[a4paper,top=3cm,bottom=2cm,left=3cm,right=3cm,marginparwidth=1.75cm]{geometry}
\usepackage{setspace}
\usepackage[table]{xcolor}
\usepackage{authblk}
\usepackage{amsmath}
\usepackage{mathrsfs}
\usepackage{amssymb}
\usepackage{enumitem}
\usepackage{graphicx}
\usepackage[colorinlistoftodos]{todonotes}
\usepackage[colorlinks=true, allcolors=blue]{hyperref}
\usepackage{wrapfig}
\usepackage{subcaption}
\usepackage{booktabs}
\usepackage{natbib}
\usepackage[linesnumbered,boxruled,lined]{algorithm2e}
\usepackage{algpseudocode}

\title{Mixtures of spatial factor analyzers for tensor-variate data}
\date{}
\author[a,*]{Hanzhang Lu}
\author[a]{Keiran Malott}
\author[a]{Kirsty Milligan}
\author[b]{Sanjeena Subedi}
\author[d]{Edana Cassol}
\author[e]{Vinita Chauhan}
\author[c]{Connor McNairn}
\author[c]{Prarthana Pasricha}
\author[c]{Sangeeta Murugkar}
\author[c]{Rowan M. Thomson}
\author[a]{Andrew Jirasek}
\author[a]{Jeffrey L. Andrews}

\affil[a]{Department of Mathematics, Physics and Statistics,
University of British Columbia, Kelowna, BC V1V 1V7, Canada}
\affil[b]{School of Mathematics and Statistics,
Carleton University, Ottawa, ON K1S 5B6, Canada}
\affil[c]{Department of Physics,
Carleton University, Ottawa, ON K1S 5B6, Canada}
\affil[d]{Department of Health Sciences,
Carleton University, Ottawa, ON K1S 5B6, Canada}
\affil[e]{Health Canada, Ottawa, ON K1A 0K9, Canada}

\affil[*]{\textit{Corresponding author:} Hanzhang Lu
(\href{mailto:hanzhang.lu@ubc.ca}{hanzhang.lu@ubc.ca})}

\begin{document}
\maketitle

\begin{abstract}
A mixture of spatial factor analyzers (MSFA) is introduced to address the challenges of clustering high-dimensional spatial data. The proposed framework utilizes the given coordinate system to implement a flexible, spline-based spatial decay covariance structure. In this approach, the number of covariance parameters increases linearly with the number of spatial locations, rather than increasing quadratically. To model non-spatial dependence, matrix variate factor analyzers are employed for further dimensionality reduction. Parameter estimation is conducted via a variant of the expectation-maximization algorithm combined with a generalized least squares estimator. The proposed models are explored in the context of tensor-variate data analysis, where simulation studies and three applications---Raman spectroscopy, SpecTex hyperspectral textures, and the Salinas hyperspectral scene---illustrate their ability to recover and distinguish spatial dependence patterns. The R implementation is available on CRAN: \url{https://CRAN.R-project.org/package=SpatMix}.

\end{abstract} 


\section{Introduction} \label{intro}

Modern data acquisition increasingly yields complex, high-dimensional datasets characterized by intricate structural dependencies. This is particularly evident in spatial statistics, where observations often take the form of matrices or tensors rather than simple vectors. In such scenarios, a single observation may represent multiple spatial systems—such as a hyperspectral image or a sensor grid—resulting in data that exhibits both local spatial correlations within the grid and global multivariate dependencies between the variables.

While model-based clustering has made significant strides in handling multi-way data \citep{viroli2011finite, dougru2016finite, gallaugher2018finite, silva2023finite}, existing multi-way models often treat dimensions efficiently but fail to explicitly account for the spatial autocorrelation inherent in spatial domains. Conversely, traditional spatial models rarely scale well to high-dimensional multivariate settings \citep{lee2025clustering} or fail to model the dependencies between replicated spatial systems \citep{lu2026spatialcovarianceconstraintsgaussian}.

To address these challenges, we propose a mixture of spatial factor analyzers model (MSFA). This framework integrates a parsimonious linear spatial covariance structure with matrix factor analysis to enable simultaneous clustering and structural inference. Specifically, we utilize a coordinate-based covariance structure to model spatial decay with a minimal number of parameters, while employing factor analyzers to capture low-dimensional patterns across the multivariate responses. We incorporate the generalized least squares (GLS) estimator \citep{browne1974generalized} with an alternating expectation-conditional maximization (AECM) algorithm \citep{meng1997algorithm} for efficient parameter estimation. To validate the proposed method, we conduct extensive simulation studies and three real-data applications involving Raman spectroscopy, the SpecTex hyperspectral texture database, and the Salinas hyperspectral scene. 

\section{Background} \label{sec:back}

\subsection{Model-based clustering} \label{sec: model-based clustering}
Model-based clustering is a statistical method that utilizes finite mixture models to infer the underlying group structure within data. The use of mixture models for clustering was first introduced by \cite{wolfe1963object} and has been extensively developed in recent decades \citep{mclachlan2000finite,mcnicholas2016mixture}. Finite mixture models assume that a heterogeneous population comprises several subpopulations, each of which can typically be represented by distributions from established families, such as Gaussian distributions \citep{wolfe1963object,celeux95,mcnicholas08}, \textit{t}-distributions \citep{andrews11,andrews12}, or generalized hyperbolic distributions \citep{browne2015mixture}. The density function of a parametric $G$-component mixture model is defined as follows:
\begin{equation} \label{eq: General FMMs}
    f(\mathbf{x}\mid \boldsymbol{\vartheta}) = \sum_{g=1}^G \pi_gf(\mathbf{x}\mid\boldsymbol{\theta_g}),
\end{equation}
where $\pi_g>0$ denotes the $g$th mixing proportion, such that $\sum_{g=1}^G\pi_g = 1$, $f(\mathbf{x}\mid\boldsymbol{\theta_g})$ represents the density of the $g$th component with parameters $\boldsymbol{\theta}_g$, and $\boldsymbol{\vartheta} = \left\{\pi_1,\dots,\pi_G,\boldsymbol{\theta}_1,\dots,\boldsymbol{\theta}_G\right\}$ represents the complete parameter space. The expectation-maximization algorithm is commonly used to fit finite mixture models \citep{dempster1977maximum,mclachlan07}. Further details regarding finite mixture models are available in works by \cite{mclachlan1988mixture,mclachlan2000finite,fruhwirth2006finite}.

\subsection{Matrix variate Gaussian distribution} \label{sec: MVG}
Analogous to the central role of the Gaussian distribution in multivariate analysis, the matrix variate Gaussian distribution is fundamental within the context of matrix variate distributions. Given a $p \times q$ location matrix $\mathbf{M}$, a $p \times p$ row covariance matrix $\boldsymbol{\Xi}$, and a $q \times q$ column covariance matrix $\boldsymbol{\Omega}$, the density of a matrix normal random matrix $\mathscr{X} \sim \mathcal{N}_{p \times q}(\mathbf{M},\boldsymbol{\Xi},\boldsymbol{\Omega})$ \citep{dawid1981some, gupta2018matrix} is given by
\begin{equation} \label{eq: MVG density}
    \phi_{p \times q}(\mathbf{X} \mid \mathbf{M},\boldsymbol{\Xi},\boldsymbol{\Omega}) = \frac{\operatorname{exp}\left\{-\frac{1}{2}\operatorname{tr}\left(\boldsymbol{\Xi}^{-1}(\mathbf{X}-\mathbf{M})\boldsymbol{\Omega}^{-1}(\mathbf{X}-\mathbf{M})^\prime\right)\right\}}{(2\pi)^{\frac{pq}{2}}{|\boldsymbol{\Xi}|^{\frac{q}{2}}|\boldsymbol{\Omega}|^{\frac{p}{2}}}}.
\end{equation}
Although allowing for matrix variate input, the matrix normal distribution remains a special case of the multivariate normal distribution. Specifically, if $\mathscr{X} \sim \mathcal{N}_{p \times q}(\mathbf{M},\boldsymbol{\Xi},\boldsymbol{\Omega})$, then $\operatorname{vec}(\mathscr{X}) \sim \mathcal{N}_{pq}(\operatorname{vec}(\mathbf{M}), \boldsymbol{\Omega}\otimes\boldsymbol{\Xi})$, where $\mathcal{N}_{pq}(\cdot)$ denotes the $pq$-dimensional multivariate normal distribution, $\operatorname{vec}(\cdot)$ is the vectorization operator, and $\otimes$ denotes the Kronecker product. Following this pattern, by introducing more covariance matrices into the Kronecker structure, higher-order inputs can be accommodated, namely the tensor-variate normal distribution \citep{ohlson2013multilinear}. In addition, as with Gaussian mixture models (GMM), a mixture of matrix normal distributions is proposed by \cite{viroli2011finite} as a straightforward extension of the matrix variate Gaussian distribution. 

It is important to note that the matrix variate Gaussian distribution introduces an identifiability issue due to the Kronecker product covariance structure. For any $a > 0$, $\mathscr{X} \sim \mathcal{N}_{p \times q}(\mathbf{M},\boldsymbol{\Xi}, \boldsymbol{\Omega})$ is equivalent to $\mathscr{X} \sim \mathcal{N}_{p \times q}(\mathbf{M},\frac{1}{a}\boldsymbol{\Xi},a\boldsymbol{\Omega})$. Hereafter, to avoid this issue, we impose the constraint $\operatorname{tr}(\boldsymbol{\Xi})=p$.

\subsection{Matrix variate factor analysis} \label{sec: MVFA}

The factor analysis model \citep{spearman04} assumes that an observed random variable is a linear combination of several latent factors, which was subsequently extended to the matrix variate context by \cite{xie2008matrix}. In matrix variate factor analysis (MVFA), a $p \times q$ random matrix $\mathscr{X}_i$ is decomposed as follows
\begin{equation} \label{eq: MVFA}
    \mathscr{X}_i = \mathbf{M}+\mathbf{L}\mathscr{U}_i\mathbf{R}^\prime + \mathscr{E}_i,
\end{equation}
where $\mathbf{M}$ is the $p \times q$ location matrix, $\mathbf{L}$ is the $p \times r$ left factor loadings matrix, $\mathbf{R}$ is the $q \times t$ right factor loading matrix, $\mathscr{U}_i \sim \mathcal{N}_{r \times t}(\mathbf{0},\mathbf{I}_r,\mathbf{I}_t)$ is the latent factor matrix, and $\mathscr{E}_i\sim \mathcal{N}_{p \times q}(\mathbf{0},\sigma^2\mathbf{I}_p,\sigma^2\mathbf{I}_q)$ is the error matrix, with $r<p$ and $t<q$. Moreover, $\mathscr{U}_i$ and $\mathscr{E}_i$ are assumed to be independent. A limitation of this specification is that $\mathscr{X}_i$ does not follow a matrix normal distribution, which reduces some of the computational advantages associated with Gaussian models.

By introducing two projected error terms into \eqref{eq: MVFA}, \cite{zhao2012bilinear} extend the model to bilinear probabilistic principal component analysis (BPPCA), which allows the random matrix $\mathscr{X}_i$ to retain normality. The resulting model can be written as
\begin{equation}\label{eq: BPPCA}
    \mathscr{X}_i = \mathbf{M}+\mathbf{L}\mathscr{U}_i\mathbf{R}^\prime + \mathbf{L}\mathscr{E}_i^{\text{R}} + \mathscr{E}_i^{\text{L}}\mathbf{R}^\prime + \mathscr{E}_i,
\end{equation}
where $\mathscr{E}_i^{\text{R}} \sim \mathcal{N}_{r \times q}(\mathbf{0},\mathbf{I}_r,\sigma^2_\text{R}\mathbf{I}_q)$, $\mathscr{E}_i^{\text{L}} \sim \mathcal{N}_{p \times t}(\mathbf{0},\sigma^2_\text{L}\mathbf{I}_p,\mathbf{I}_t)$, and $\mathscr{E}_i \sim \mathcal{N}_{p \times q}(\mathbf{0},\sigma^2_\text{L}\mathbf{I}_p,\sigma^2_\text{R}\mathbf{I}_q)$. The remaining terms in \eqref{eq: BPPCA} are defined as in \eqref{eq: MVFA}. Moreover, $\mathscr{U}_i$, $\mathscr{E}_i^{\text{R}}$, $\mathscr{E}_i^{\text{L}}$, and $\mathscr{E}_i$ are independent of each other. Under \eqref{eq: BPPCA}, $\mathscr{X}_i$ has the distribution 
\begin{equation*}
    \mathscr{X}_i \sim \mathcal{N}_{p \times q}(\mathbf{M},\mathbf{L}\mathbf{L}^\prime+\sigma^2_\text{L}\mathbf{I}_p,\mathbf{R}\mathbf{R}^\prime+\sigma^2_\text{R}\mathbf{I}_q).
\end{equation*}

In a manner analogous to the mixture of factor analyzers \citep{ghahramani1996algorithm}, BPPCA is further extended to a mixture of matrix variate bilinear factor analyzers (MMVBFA) by \citet{gallaugher2018proceedings}. A distinct extension to skewed matrix variate distributions was developed by \citet{gallaugher2020mixtures}. Taking a different approach, \cite{sharp2023dual} propose another parsimonious matrix normal mixture by a dual-subspace projection that constrains the trailing eigenvalues of the covariance matrices.

\subsection{Spatially constrained Gaussian mixture models} \label{sec: spatial constraints}

\cite{worsley1991linear} proposed a linear spatial correlation model to characterize isotropic spatial autocorrelation within observations. Consider an $M$-order random tensor $\mathcal{X} \in \mathbb{R}^{p_1 \times p_2 \times \cdots \times p_M}$. We denote its vectorization as $\mathbf{X} = \operatorname{vec}(\mathcal{X}) \in \mathbb{R}^{p}$, where $p = \prod_{m=1}^M p_m$. The operator $\operatorname{vec}(\cdot)$ transforms the tensor $\mathcal{X}$ into a vector by mapping the element with the multi-index $(i_1, \dots, i_M)$ to the $j$-th entry of $\mathbf{X}$, according to the mapping $j = 1 + \sum_{m=1}^M (i_m - 1) \prod_{m^\prime = 1}^{m-1} p_{m^\prime}$. Crucially, each element $X_j$ (for $j=1, \dots, p$) in the vectorized form is associated with a unique coordinate vector $\mathbf{c}_j$, which stands in a one-to-one correspondence with the original tensor subscript $(i_1, \dots, i_M)$.
%Some specific coordinate systems $\mathcal{C} = {\mathbf{c}_1, \dots, \mathbf{c}_p}$ induce a natural mapping that reshapes the vector $\mathbf{X}$ into a tensor form $\mathscr{X}$, where the coordinate vectors $\mathbf{c}_i, i=1,\dots,p$ and the subscripts of $\mathscr{X}$ are in a one-to-one correspondence. Correspondingly, we can also decompose a random tensor $\mathscr{X}$ as a vector $\mathbf{X}$ in the opposite way. 
In this context, the spatial proximity between two elements $X_i$ and $X_j$ is quantified by the Euclidean distance $d_{ij} = \|\mathbf{c}_i - \mathbf{c}_j\|_2$. According to the linear spatial correlation model, the covariance matrix of $\mathbf{X}$ is defined as
\begin{equation} \label{eq: quadratic decay}
    \boldsymbol{\Xi} = \alpha_{1} \mathbf{J} - \alpha_{2} \mathbf{D} + \alpha_{3} \mathbf{I},
\end{equation}
where $\alpha_{1}, \alpha_{2}, \alpha_{3} \in \mathbb{R}_+$ are linear spatial parameters, $\mathbf{J}$ is a matrix of ones, $\mathbf{I}$ is the identity matrix, and $\mathbf{D} = [d_{ij}^2]$ denotes the matrix of squared Euclidean distances. This structure implies that covariance decreases quadratically as distance increases, referred to as quadratic decay (QD). A key advantage of this model is its parsimony. Since all three matrices $\mathbf{J}$, $\mathbf{D}$, and $\mathbf{I}$ are fixed, the covariance structure is fully specified by only three parameters, regardless of the data dimensionality. However, the model faces practical limitations. First, it is susceptible to multicollinearity among the basis matrices in high-dimensional spatial systems. Second, the raw Euclidean distances often differ in magnitude across coordinate systems, necessitating a manual, and often non-trivial, normalization step.

As an extension of the QD framework, \citet{lu2026spatialcovarianceconstraintsgaussian} introduced a Sigmoid Decay (SD) spatial constraint within GMM. This approach assumes that spatial covariance diminishes with distance according to a flexible sigmoid function. Similar to the QD model, the SD covariance is expressed as a weighted linear combination of the basis matrices $\mathbf{J}$, $\mathbf{D}$, and $\mathbf{I}$. However, the elements of the distance matrix $\mathbf{D}$ are transformed via the function
\begin{equation}
    D_{ij} = h(d_{ij}) = a\left( \frac{1}{1+e^{-\beta d_{ij} + 3}} - s \right),
\end{equation}
where $\beta$ is a tuning parameter controlling the rate of spatial decay. The constants $s$ and $a$ are normalization factors defined as $s = (1+e^{3})^{-1}$ and $a = [ (1+e^{-2\beta+3})^{-1} - s ]^{-1}$. This parameterization is designed such that the curve originates at $(0,0)$ and passes through $(2,1)$, thereby deterministically normalizing the relevant spatial distance to the interval $[0,2]$. Furthermore, the inclusion of the steepness parameter $\beta$ enhances model flexibility and significantly mitigates the collinearity issues inherent in the QD formulation.

\section{Methodology}\label{method}
Consider $q$ spatial systems defined over a common coordinate system $\mathcal{C} = \{\mathbf{c}_1, \dots, \mathbf{c}_p\}$. Let $\mathscr{X}$ denote a $p \times q$ random matrix where the $j$-th column, $\mathbf{X}_j = \operatorname{vec}(\mathcal{X}_j)$, represents the vectorized observations of the $j$-th spatial system ($j=1,\dots,q$). Under this formulation, the $i$-th row of $\mathscr{X}$ corresponds to the observations associated with the spatial coordinate $\mathbf{c}_i$ across all $q$ systems.

\subsection{Flexible spatial decay covariance structure} \label{sec: FSD}
\begin{figure}[ht]
    \centering
    \includegraphics[width=1\linewidth]{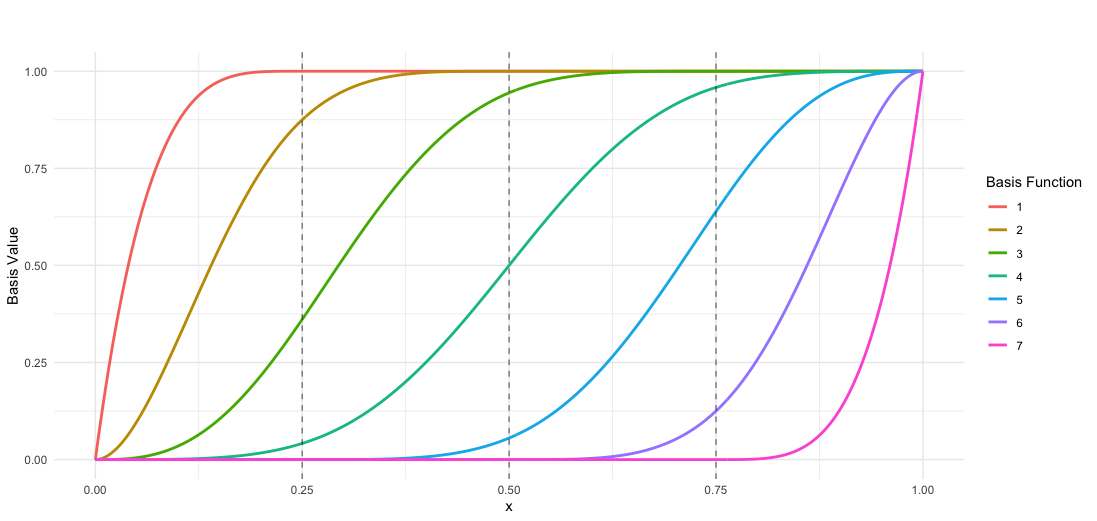}
    \caption{Visualization of cubic I-spline basis functions defined on the interval [0, 1]. The vertical dashed lines represent the internal knots located at 0.25, 0.5, and 0.75.}
    \label{fig: I-splines Viz}
\end{figure}

As previously discussed, the SD structure is more flexible and relaxes the collinearity issue present in the QD by introducing a parametric sigmoid function. However, the sigmoid relationship may not always be appropriate, and it also constrains the variance of all elements in a spatial system to be the same. Moreover, the domain of $\beta_g$ to guarantee the positive definiteness of the covariance matrix remains unclear. This work introduces a flexible spatial decay (FSD) covariance structure, which assumes, in general, that the isotropic spatial correlation decreases as the distance increases. Rather than relying on a predefined function to model the decay pattern, the FSD structure employs I-splines \citep{ramsay1988monotone} to model the decay pattern nonparametrically. As integrals of non-negative M-spline \citep{curry1966polya} basis functions, I-spline basis functions are inherently non-negative and non-decreasing, as shown in Figure~\ref{fig: I-splines Viz}, making them suitable for modeling monotonic relationships. Furthermore, normalizing distances from any coordinate system to the interval $[0,1]$ eliminates the need for manual adjustment. Over this normalized domain, ordinary I-splines constrained by probability simplex coefficients can flexibly fit non-decreasing curves that strictly begin at the origin $(0,0)$ and terminate at $(1,1)$.
Given a series of internal knots $\mathbf{t} = (t_1, \dots, t_{k})$ with $t_1 > 0$ and $t_k < 1$, and degree $m$, let $\mathcal{I}_j(x\mid \mathbf{t},m)$ denote the $j$th I-spline basis function. The decay function is defined as
\begin{equation}
    h(x\mid \mathbf{t},m) = \sum_{j=1}^{k+m+1}\beta_j\mathcal{I}_j(x\mid \mathbf{t},m),
\end{equation}
where $\boldsymbol{\beta} = (\beta_{1},\beta_{2},\dots,\beta_{k+m+1})$ is a probability simplex, such that $\beta_{j} \geq 0$ and $\sum_{j=1}^{k+m+1} \beta_{j} = 1$. The flexible spatial covariance matrix can be expressed as 
\begin{equation} \label{eq: flexible spatial covariance}
    \boldsymbol{\Xi} = \alpha_{1} \mathbf{J} - \alpha_{2} \mathbf{D}(\boldsymbol{\beta}\mid \mathbf{t},m) + (\boldsymbol{\gamma}\mathbf{1}^\prime) \circ \mathbf{I},
\end{equation}
where $\mathbf{J}$, $\mathbf{I}$, $\alpha_1$ and $\alpha_2$ retain their definitions from the previous section, $\mathbf{D}(\boldsymbol{\beta}\mid \mathbf{t},m)$ has the $(i,j)$ entry $h(d_{ij} \mid \mathbf{t},m)$, $\boldsymbol{\gamma} = (\alpha_3,\dots,\alpha_{p+2})^\prime \in \mathbb{R}_+^p$ is a $p$-dimensional vector, $\mathbf{1}$ is a vector of ones, and $\circ$ is Hadamard product. Note that $(\boldsymbol{\gamma}\mathbf{1}^\prime) \circ \mathbf{I} = \operatorname{diag}(\boldsymbol{\gamma})$, which is a diagonal matrix with elements $\{\alpha_3,\dots,\alpha_{p+2}\}$. Compared to the QD and SD structures, \eqref{eq: flexible spatial covariance} relaxes the limitations
on both the shape of the decay curve and the variance. As a special case, setting $\boldsymbol{\gamma} = \alpha_3 \mathbf{1}$ reduces the noise term $(\alpha_3 \mathbf{1}\mathbf{1}^\prime) \circ \mathbf{I}$ to $\alpha_3 \mathbf{I}$, recovering the homoscedastic structure assumed by the QD and SD models.
% Compared to the QD and SD, \eqref{eq: flexible spatial covariance} relaxes the limitations not only on the shape of the decay curve but also on the variance. Conversely, by setting $\boldsymbol{\gamma} = \alpha_3 \mathbf{1}$, it reduces to the constrained form $(\alpha_3 \mathbf{1}\mathbf{1}^\prime) \circ \mathbf{I} = \alpha_3 \mathbf{I}$. 
Given a fixed knot count $k$ and spline degree $m$, the spatial covariance in \eqref{eq: flexible spatial covariance} includes $p+k+m+1$ free parameters. These parameters consist of $p$ location-specific diagonal variances, two scale parameters, and $k+m$ independent spline coefficients determined by the simplex sum-to-one constraint. Consequently, the number of parameters increases linearly with $p$, as opposed to quadratically. If a common diagonal variance is assumed, the total parameter count becomes $k+m+1$, which does not depend on $p$. Furthermore, following the guidance of \citet{ramsay1988monotone}, we place internal knots at empirical quantiles of the pairwise spatial distances. The author also notes that relatively few interior knots and a low spline order are often sufficient in statistical applications, with greater flexibility used when warranted by the data. Moreover, on a regular Cartesian grid, $\boldsymbol{\Xi}$ is a multilevel Toeplitz matrix plus a diagonal term. As a result, multidimensional fast Fourier transforms (FFTs) can be employed to accelerate matrix–vector products with $\boldsymbol{\Xi}$, which may facilitate iterative solutions of spatial covariance systems \citep{chen2014parallel}.

For parameter estimation, a nested two-stage optimization procedure is conducted, incorporating the generalized least squares (GLS) estimator \citep{browne1974generalized}. Let $\boldsymbol{\alpha} = (\alpha_1, \alpha_2, \boldsymbol{\gamma}^\prime)^\prime$, where $\boldsymbol{\gamma} = (\alpha_3, \dots, \alpha_{p+2})^\prime$. In the first stage, $\boldsymbol{\alpha}$ is estimated by minimizing 
\begin{equation}\label{eq: glse_g}
    g(\boldsymbol{\alpha}) = \frac{1}{2} (\mathbf{s}-\boldsymbol{\Delta}\boldsymbol{\alpha})^\prime(\mathbf{V}\otimes\mathbf{V})(\mathbf{s}-\boldsymbol{\Delta}\boldsymbol{\alpha}),
\end{equation}
where $\mathbf{s}$ is the vectorization of the sample covariance matrix $\mathbf{S}$, $\mathbf{V}$ is a positive definite weight matrix, and $\boldsymbol{\Delta}$ is the design matrix. The first two columns of $\boldsymbol{\Delta}$ are defined as $\operatorname{vec}(\mathbf{J})$ and $-\operatorname{vec}(\mathbf{D}(\boldsymbol{\beta}\mid \mathbf{t},m))$, respectively. For $i > 2$, the $i$-th column corresponds to the parameter $\alpha_i$ and is defined as $\operatorname{vec}(\mathbf{E}_{i-2})$, where $\mathbf{E}_{k}$ is a matrix with 1 at the $(k, k)$ entry and 0 elsewhere. Setting the first derivative of \eqref{eq: glse_g} equal to zero yields
\begin{equation} %\label{eq: alpha g}
    \hat{\boldsymbol{\alpha}} = \left\{\boldsymbol{\Delta}^\prime(\mathbf{V}\otimes \mathbf{V})\boldsymbol{\Delta}\right\}^{-1}\boldsymbol{\Delta}^\prime\operatorname{vec}(\mathbf{V}\mathbf{S}\mathbf{V}).
\end{equation}
Note that when the homoscedastic noise constraint $\boldsymbol{\gamma} = \alpha_3 \mathbf{1}$ is imposed, $\boldsymbol{\Delta}$ reduces to three columns: $\operatorname{vec}(\mathbf{J})$, $-\operatorname{vec}(\mathbf{D}(\boldsymbol{\beta}\mid \mathbf{t},m))$, and $\operatorname{vec}(\mathbf{I})$.

In the second stage, we partition the estimate $\hat{\boldsymbol{\alpha}}$ into its components $\hat{\alpha}_1$, $\hat{\alpha}_2$, and the noise parameter vector $\hat{\boldsymbol{\gamma}}$. Let $\mathscr{I}$ denote the I-spline design matrix, where the $j$-th column corresponds to $\mathcal{I}(x\mid \mathbf{t}, m)$ evaluated at the observed distances in the spatial systems. Given the estimates from the first stage, we estimate the coefficients $\boldsymbol{\beta}$ by minimizing the following GLS objective function
\begin{equation}\label{eq: glse_g_beta}
    g(\boldsymbol{\beta}) = \frac{1}{2} (\mathbf{r} -  \mathscr{I}\boldsymbol{\beta})^\prime(\mathbf{V}\otimes\mathbf{V})(\mathbf{r} - \mathscr{I}\boldsymbol{\beta}),
\end{equation}
on the probability simplex, where $\mathbf{r} = \hat{\alpha}_{2}^{-1}\operatorname{vec}\left(\hat{\alpha}_{1} \mathbf{J} + (\hat{\boldsymbol{\gamma}}\mathbf{1}^\prime) \circ \mathbf{I} - \mathbf{S}\right)$, assuming $\hat{\alpha}_2 \ne 0$. Since $\mathbf{V}$ is positive definite, the objective function in \eqref{eq: glse_g_beta} is convex. Because $\boldsymbol{\beta}$ is constrained to the probability simplex, we employ the Projected Gradient Descent (PGD) algorithm \citep{bauschke2023introduction}. The steepest descent direction is given by the negative gradient
\begin{equation} \label{eq: direction}
    -\nabla g(\boldsymbol{\beta}) = \mathscr{I}^\prime (\mathbf{V}\otimes\mathbf{V})(\mathbf{r} - \mathscr{I}\boldsymbol{\beta}).
\end{equation}
Let $\boldsymbol{\beta}^*$ denote the candidate update obtained by taking a step in the direction of \eqref{eq: direction}. To project $\boldsymbol{\beta}^*$ onto the probability simplex, according to the work by \cite{duchi2008efficient}, we first sort its elements in descending order to obtain $\mathbf{u}=(u_1, \dots, u_{K})$, where $K = k+m+1$. We then compute $\rho = \operatorname{max}\{j : u_j + \frac{1}{j}(1-\sum_{i=1}^j u_i) > 0\}$ and define the threshold $\lambda = \frac{1}{\rho}(1-\sum_{i=1}^\rho u_i)$. The projected elements are updated as ${\beta}_{i} = \operatorname{max}\{\beta^*_{i}+\lambda, 0\}$. Starting from an initial $\boldsymbol{\beta}_0$, we iteratively perform the gradient descent step and the simplex projection until a convergence criterion is satisfied. To accelerate convergence, we utilize the Barzilai-Borwein (BB) step-size method \citep{barzilai1988two}.  Under the regularity conditions of \cite{browne1974generalized}, a consistent GLS covariance estimator converges in probability to the true positive definite covariance matrix, with \(p\) fixed. Thus, $\hat{\boldsymbol{\Xi}}$ is positive definite with probability tending to one.

\subsection{Mixtures of spatial factor analyzers}

Consider $N$ independent and identically distributed (i.i.d.) random matrices $\mathscr{X}_1, \dots, \mathscr{X}_N$, each of size $p \times q$, representing $q$ dependent and vectorized spatial systems that share a common coordinate system $\mathcal{C}$. Building on the MVFA model introduced in Section~\ref{sec: MVFA}, if $\mathscr{X}_i$ follows a mixture of spatial factor analyzers, specifically, with probability $\pi_g$, $\mathscr{X}_i$ is generated by
\begin{equation} \label{eq: MSFA}
    \mathscr{X}_i = \mathbf{M}_g + \mathscr{U}_{ig} \boldsymbol{\Lambda}_g^{\prime} + \mathscr{E}_{ig},
\end{equation}
where $\mathbf{M}_g$ is a $p \times q$ location matrix, $\boldsymbol{\Lambda}_g$ is a $q \times r$ matrix of column factor loadings, $\mathscr{U}_{ig} \sim \mathcal{N}_{p \times r}\left(\mathbf{0}, \boldsymbol{\Xi}_g, \mathbf{I}_r\right)$ is the spatial factor matrix, and $\mathscr{E}_{ig} \sim \mathcal{N}_{p \times q}\left(\mathbf{0}, \boldsymbol{\Xi}_g, \boldsymbol{\Psi}_g\right)$. The latent spatial factor matrix $\mathscr{U}_{ig}$ and the error term $\mathscr{E}_{ig}$ are assumed to be independent of each other. Here, the spatial covariance matrix is defined as $\boldsymbol{\Xi}_g = \alpha_{1g} \mathbf{J} - \alpha_{2g} \mathbf{D}(\boldsymbol{\beta}_g \mid  \mathbf{t},m) + (\boldsymbol{\gamma}_g\mathbf{1}^\prime) \circ \mathbf{I}$, following the structure defined in Section~\ref{sec: FSD}. $\boldsymbol{\Psi}_g = \operatorname{diag}(\psi_{1g},\psi_{2g},\dots,\psi_{qg})$, with $\psi_{jg}\in \mathbb{R}_+$ for $j=1,\ldots,q$, is the diagonal uniqueness matrix. 

Given the latent component membership $\mathbf{z}_i =(z_{i1},\dots,z_{iG})$, where $z_{ig}=1$ indicates that $\mathscr{X}_i$ belongs to component $g$, under \eqref{eq: MSFA}, the conditional distribution of $\mathscr{X}_i$ is
\begin{equation*}
    \mathscr{X}_i \mid z_{ig} = 1 \sim \mathcal{N}_{p \times q} (\mathbf{M}_g, \alpha_{1g} \mathbf{J} - \alpha_{2g} \mathbf{D}(\boldsymbol{\beta}_g \mid  \mathbf{t},m) + (\boldsymbol{\gamma}_g\mathbf{1}^\prime) \circ \mathbf{I}, \boldsymbol{\Lambda}_g\boldsymbol{\Lambda}_g ^{\prime} + \boldsymbol{\Psi}_g).
\end{equation*}
Consequently, the density of $\mathscr{X}_i$ can be written as
\begin{equation*}
    f(\mathbf{X}\mid \boldsymbol{\vartheta}) = \sum_{g=1}^G \pi_g \phi_{p \times q}\left(\mathbf{X} \mid \mathbf{M}_g, \boldsymbol{\Xi}_g, \boldsymbol{\Omega}_g\right),
\end{equation*}
where $\boldsymbol{\Omega}_g = \boldsymbol{\Lambda}_g\boldsymbol{\Lambda}_g^{\prime}+\boldsymbol{\Psi}_g$, $\phi_{p \times q}(\cdot)$ refers to the $p \times q$-dimensional matrix variate normal density function defined in \eqref{eq: MVG density}.

\subsection{Parameter estimation}

Given a set of $N$ observations denoted by $\mathbf{X}_1, \dots, \mathbf{X}_N$, the observed log-likelihood function is given by
\begin{equation*}
    \ell(\boldsymbol{\vartheta})=\sum_{i=1}^N \log \sum_{g=1}^G \pi_g \phi_{p \times q}\left(\mathbf{X}_i \mid \mathbf{M}_g,  \boldsymbol{\Xi}_g, \boldsymbol{\Omega}_g\right),
\end{equation*}
where $\boldsymbol{\Xi}_g = \alpha_{1g} \mathbf{J} - \alpha_{2g} \mathbf{D}(\boldsymbol{\beta}_g \mid  \mathbf{t},m) + (\boldsymbol{\gamma}_g\mathbf{1}^\prime) \circ \mathbf{I}$ represents the spatial covariance, and $\boldsymbol{\Omega}_g = \boldsymbol{\Lambda}_g\boldsymbol{\Lambda}_g^{\prime}+\boldsymbol{\Psi}_g$ represents the across-system covariance.

The proposed model incorporates two sets of latent variables: the latent component membership $\mathbf{z}_i$ and the latent spatial factors $\mathscr{U}_{ig}$. Parameter estimation is performed using a three-stage alternating expectation-conditional maximization (AECM) algorithm \citep{meng1997algorithm}, which integrates the nested optimization procedure described in Section~\ref{sec: FSD}.

\paragraph{Stage 1:} In the first stage, the mixing proportions $\boldsymbol{\pi}_g$, and the location matrix $\mathbf{M}_g$ are estimated conditional on the estimated component covariance matrices $\hat{\boldsymbol{\Xi}}_g$ and $\hat{\boldsymbol{\Omega}}_g$. Hence, the conditional expectation of the complete log-likelihood is 
\begin{equation*}
    \ell_c^{(1)}=C_1+\sum_{g=1}^G \sum_{i=1}^N \hat{z}_{i g}\left\{\log \pi_g-\frac{1}{2} \operatorname{tr}\left\{\hat{\boldsymbol{\Xi}}_g^{-1}\left(\mathbf{X}_i-\mathbf{M}_g\right) \hat{\boldsymbol{\Omega}}_g^{-1}\left(\mathbf{X}_i-\mathbf{M}_g\right)^{\prime}\right\}\right\},
\end{equation*}
where $C_1$ is the constant with respect to $\pi_g$ and $\mathbf{M}_g$ and $\hat{z}_{ig}$ is the estimated latent membership, which is updated in E-step via
\begin{equation*}
    \hat{z}_{i g}=\frac{\pi_g \phi_{p \times q}\left(\mathbf{X}_i \mid \hat{\mathbf{M}}_g, \hat{\boldsymbol{\Xi}}_g, \hat{\boldsymbol{\Omega}}_g\right)}{\sum_{h=1}^G \pi_h \phi_{p \times q}\left(\mathbf{X}_i \mid \hat{\mathbf{M}}_h, \hat{\boldsymbol{\Xi}}_h, \hat{\boldsymbol{\Omega}}_h\right)}.
\end{equation*}
Then, the update equations for $\pi_g$ and $\mathbf{M}_g$ are given by
\begin{equation*}
    \hat{\pi}_g=\frac{N_g}{N} \quad \text { and } \quad \hat{\mathbf{M}}_g=\frac{1}{N_g} \sum_{i=1}^N \hat{z}_{i g} \mathbf{X}_i,
\end{equation*}
where $N_g=\sum_{i=1}^N \hat{z}_{i g}$, for $g=1,\dots,G$.

\paragraph{Stage 2:} In the second stage, we estimate the spatial covariance parameters $\alpha_{1g}, \alpha_{2g}, \boldsymbol{\gamma}_g$, and $\boldsymbol{\beta}_g$, conditional on the current estimates of $\hat{\pi}_g$, $\hat{\mathbf{M}}_g$, and $\hat{\boldsymbol{\Omega}}_g$. The marginal log-likelihood for this stage is given by 
\begin{equation} \label{eq: single system loglikelihood2}
    \begin{aligned}
        \ell_c^{(2)}(\boldsymbol{\vartheta}) = C_2 &- \frac{1}{2}\sum_{g=1}^G qN_g\log|\boldsymbol{\Xi}_g| - \frac{1}{2} \sum_{g=1}^G qN_g \operatorname{tr}\left\{\boldsymbol{\Xi}_g^{-1}\mathbf{S}_g\right\} \\
        = C_2 &- \frac{1}{2}\sum_{g=1}^G qN_g\log|\alpha_{1g}\mathbf{J} - \alpha_{2g}\mathbf{D}_g(\boldsymbol{\beta}_g \mid  \mathbf{t},m) + (\boldsymbol{\gamma}_g\mathbf{1}^\prime) \circ \mathbf{I}| \\
        &  - \frac{1}{2} \sum_{g=1}^G qN_g \operatorname{tr}\left\{
        (\alpha_{1g}\mathbf{J} - \alpha_{2g}\mathbf{D}_g(\boldsymbol{\beta}_g \mid  \mathbf{t},m) + (\boldsymbol{\gamma}_g\mathbf{1}^\prime) \circ \mathbf{I})^{-1}\mathbf{S}_g\right\},
    \end{aligned}
\end{equation}
where $C_2$ is a constant independent of the spatial parameters, the group-specific spatial sample covariance matrix $\mathbf{S}_g$ is defined as
\begin{equation*}
    \mathbf{S}_g = \frac{1}{qN_g}\sum_{i=1}^N \hat{z}_{ig}\left(\mathbf{X}_i-\mathbf{M}_g\right) \hat{\boldsymbol{\Omega}}_g^{-1}\left(\mathbf{X}_i-\mathbf{M}_g\right)^\prime.
\end{equation*}
Following the theory established by \cite{browne1974generalized}, maximizing \eqref{eq: single system loglikelihood2} is asymptotically equivalent to minimizing a GLS objective function, so a group-specific estimation procedure as in Section~\ref{sec: FSD} is used. We set the weight matrix $\mathbf{V}^*_g$ to be the inverse of the estimated spatial covariance from the previous EM iteration, denoted as $\mathbf{V}^*_g = (\hat{\boldsymbol{\Xi}}^*_g)^{-1}$.

 First, given the current spline coefficients $\hat{\boldsymbol{\beta}}_g$, the linear parameters $\boldsymbol{\alpha}_g = (\alpha_{1g}, \alpha_{2g}, \boldsymbol{\gamma}_g^\prime)^\prime$ are updated via
\begin{equation} %\label{eq: alpha g}
    \hat{\boldsymbol{\alpha}}_g = \left\{\boldsymbol{\Delta}_g^\prime(\mathbf{V}_g^*\otimes \mathbf{V}_g^*)\boldsymbol{\Delta}_g\right\}^{-1}\boldsymbol{\Delta}_g^\prime\operatorname{vec}(\mathbf{V}_g^*\mathbf{S}_g\mathbf{V}_g^*), 
\end{equation}
where $\boldsymbol{\Delta}_g$ follows the design matrix definition in Section~\ref{sec: FSD}. Subsequently, when $\hat{\alpha}_{2g} \ne 0$, using the updated $\hat{\boldsymbol{\alpha}}_g$, we estimate $\boldsymbol{\beta}_g$ by employing the PGD algorithm to minimize
\begin{equation}%\label{eq: glse_g_beta*}
    g(\boldsymbol{\beta}) = \frac{1}{2} (\mathbf{r}_g -  \mathscr{I}\boldsymbol{\beta})^\prime(\mathbf{V}_g^*\otimes\mathbf{V}_g^*)(\mathbf{r}_g - \mathscr{I}\boldsymbol{\beta}),
\end{equation}
where $\mathbf{r}_g = \hat{\alpha}_{2g}^{-1}\operatorname{vec}(\hat{\alpha}_{1g} \mathbf{J} + (\hat{\boldsymbol{\gamma}_g}\mathbf{1}^\prime) \circ \mathbf{I} - \mathbf{S}_g)$. The PGD iterations continue until the log-likelihood function no longer improves, ensuring the solution satisfies the probability simplex constraint.

\paragraph{Stage 3:} In the final stage, the complete data consists of the observations $\mathbf{X}_1, \dots, \mathbf{X}_N$, the latent component memberships $\mathbf{z}_1,\dots,\mathbf{z}_N$, and the latent spatial factors $\mathscr{U}_{ig}$ for $i=1,\dots, N$ and $g=1,\dots, G$. The complete-data log-likelihood is given by
\begin{equation*}
    \begin{aligned}
    \ell_c^{(3)} =\; & C_3 {- \frac{p}{2}\sum_{g=1}^G N_g\log\left|\boldsymbol{\Psi}_g\right|}
    - \frac{1}{2}\sum_{g=1}^G \sum_{i=1}^N \hat{z}_{ig}\operatorname{tr}\Bigl[
    \mathbf{\Psi}_g^{-1}\bigl(\mathbf{X}_i-\hat{\mathbf{M}}_g\bigr)^{\prime}\hat{\boldsymbol{\Xi}}_g^{-1}\bigl(\mathbf{X}_i-\hat{\mathbf{M}}_g\bigr) \\[1ex]
    & {}- \boldsymbol{\Psi}_g^{-1}\boldsymbol{\Lambda}_g\,\mathscr{U}_{ig}^{\prime}\hat{\boldsymbol{\Xi}}_g^{-1}\bigl(\mathbf{X}_i-\hat{\mathbf{M}}_g\bigr)
    - \boldsymbol{\Psi}_g^{-1}\bigl(\mathbf{X}_i-\hat{\mathbf{M}}_g\bigr)^{\prime}\hat{\boldsymbol{\Xi}}_g^{-1}\mathscr{U}_{ig}\boldsymbol{\Lambda}_g^{\prime} \\[1ex]
    & {}+ \boldsymbol{\Psi}_g^{-1}\boldsymbol{\Lambda}_g\,\mathscr{U}_{ig}^{\prime}\hat{\boldsymbol{\Xi}}_g^{-1}\mathscr{U}_{ig}\boldsymbol{\Lambda}_g^{\prime}
    \Bigr],
    \end{aligned}
\end{equation*}
where $\hat{\boldsymbol{\Xi}}_g = \hat{\alpha}_{1g} \mathbf{J} - \hat{\alpha}_{2g} \mathbf{D}(\hat{\boldsymbol{\beta}}_g \mid  \mathbf{t},m) + (\hat{\boldsymbol{\gamma}}_g\mathbf{1}^\prime) \circ \mathbf{I}$, and $C_3$ collects constants independent of $\boldsymbol{\Psi}_g$ and $\boldsymbol{\Lambda}_g$. 

In the E-step, we compute the required conditional expectations based on the current parameter estimates
\begin{align*}
  u_{ig} &:= \mathrm{E}\Bigl[\mathscr{U}_{ig} \mid \mathbf{X}_i, z_{ig}=1\Bigr] = \left(\mathbf{X}_i-\mathbf{M}_g\right) \boldsymbol{\Psi}_g^{-1} \boldsymbol{\Lambda}_g 
    \left(\mathbf{I}_r+\boldsymbol{\Lambda}_g^{\prime}\boldsymbol{\Psi}_g^{-1}\boldsymbol{\Lambda}_g\right)^{-1},\\[1ex]
  v_{ig} &:= \mathrm{E}\Bigl[\mathscr{U}_{ig}^{\prime} \hat{\boldsymbol{\Xi}}_g^{-1} \mathscr{U}_{ig} \mid \mathbf{X}_i, z_{ig}=1\Bigr] = p\left(\mathbf{I}_r+\boldsymbol{\Lambda}_g^{\prime}\boldsymbol{\Psi}_g^{-1}\boldsymbol{\Lambda}_g\right)^{-1} 
         + u_{ig}^{\prime} \hat{\boldsymbol{\Xi}}_g^{-1} u_{ig}.
\end{align*}
In the CM-step, the factor loadings $\boldsymbol{\Lambda}_g$ and the column noise variance matrix $\boldsymbol{\Psi}_g$ are updated as follows
\begin{equation*}
    {\hat{\boldsymbol{\Lambda}}_g=\left[\sum_{i=1}^N \hat{z}_{ig}\left(\mathbf{X}_i-\hat{\mathbf{M}}_g\right)^{\prime}\hat{\boldsymbol{\Xi}}_g^{-1}u_{ig}\right]\left[\sum_{i=1}^N \hat{z}_{ig}v_{ig}\right]^{-1},}\qquad \hat{\boldsymbol{\Psi}}_g=\frac{1}{N_g p} \operatorname{diag}\left\{\hat{\mathbf{P}}_g \right\},
\end{equation*}
where the residual sum of squares matrix $\hat{\mathbf{P}}_g$ is given by
\begin{equation*}
    \hat{\mathbf{P}}_g=\sum_{i=1}^N \hat{z}_{i g}\left[\left(\mathbf{X}_i-\hat{\mathbf{M}}_g\right)^{\prime} \hat{\boldsymbol{\Xi}}_g^{-1}\left(\mathbf{X}_i-\hat{\mathbf{M}}_g\right)-\hat{\boldsymbol{\Lambda}}_g u_{i g}^\prime \hat{\boldsymbol{\Xi}}_g^{-1}\left(\mathbf{X}_i-\hat{\mathbf{M}}_g\right)\right].
\end{equation*}

\section{Simulation Studies} \label{sec: Simulation Studies}
We conducted four simulation studies to evaluate the practical performance of MSFA. Simulation I examines clustering and covariance recovery. Simulation II investigates statistical and computational behavior as spatial dimension increases. Simulation III examines selection of the number of components and factors. Simulation IV compares MSFA with MCLUST \citep{scrucca2016mclust}, PGMM \citep{mcnicholas08}, and MatrixMixtures \citep{tomarchio2022mixtures}.

Throughout these simulations and the subsequent real-data applications, we utilize the intrinsic matrix element subscripts as the spatial coordinate system, although the framework readily accommodates user-defined coordinates. Classification accuracy is assessed using the Adjusted Rand Index \citep[ARI;][]{hubert1985comparing}, which improves upon the standard Rand Index \citep[RI;][]{rand1971objective} by correcting for chance agreement. Furthermore, we use the Bayesian information criterion (BIC) \citep{schwarz1978estimating} to evaluate model fit, selecting the model with the largest BIC under the convention $2\ell-w\log N$, where $ w$ is the fitted parameter count.

% (coordinate system, normalization of distance, precision of location matrix, different sample sizes, BIC, number of groups, number of factors)

\subsection{Simulation I - parameter recovery} \label{sec: sim1}
In the first simulation, we demonstrate the capacity of the proposed model to recover spatial covariance structures when data are generated from distinct spatial decay functions. The spatial coordinates are arranged on a $5 \times 5 \times 5$ cubic grid, giving $p=125$ spatial locations, and the non-spatial dimension is set to $q=10$, resulting in the dimensionality $1{,}250$. After matricization, the dimensions of the resulting matrices are $125 \times 10$. The data are generated from a three-component MSFA model with mixing proportions $\boldsymbol{\pi} = (0.5, 0.3, 0.2)$. To isolate covariance recovery, all component mean matrices $\mathbf{M}_g$ are set to zero. For each component, the column covariance follows a three-factor structure, where the factor loadings are generated independently from $\mathrm{Unif}(-1,1)$ and the uniquenesses from $\mathrm{Unif}(0,1)$. The spatial row covariance matrices are generated on the normalized pairwise distance scale using three decay profiles: a quadratic decay \citep{worsley1991linear} and two sigmoid decays \citep{lu2026spatialcovarianceconstraintsgaussian} with steepness parameters $\beta=10$ and $\beta=30$. Specifically, for component $g$, the row covariance is constructed as $\mathbf{J}-\mathbf{D}_g+(\boldsymbol{\gamma}_g\mathbf{1}^\prime) \circ \mathbf{I}$, where $\mathbf{D}_g$ denotes the component-specific decay function evaluated at the normalized distances, and the diagonal noise terms are generated independently from $\mathrm{Unif}(2,4)$. In total, $100$ independent datasets are simulated, each containing $N=1{,}000$ observations. Each dataset is fitted using the correctly specified number of components and latent factors, with initialization based on covariance. The spatial covariance functions are estimated using cubic I-splines with $6$ knots. 

\begin{figure}
    \centering
    \includegraphics[width=1\linewidth]{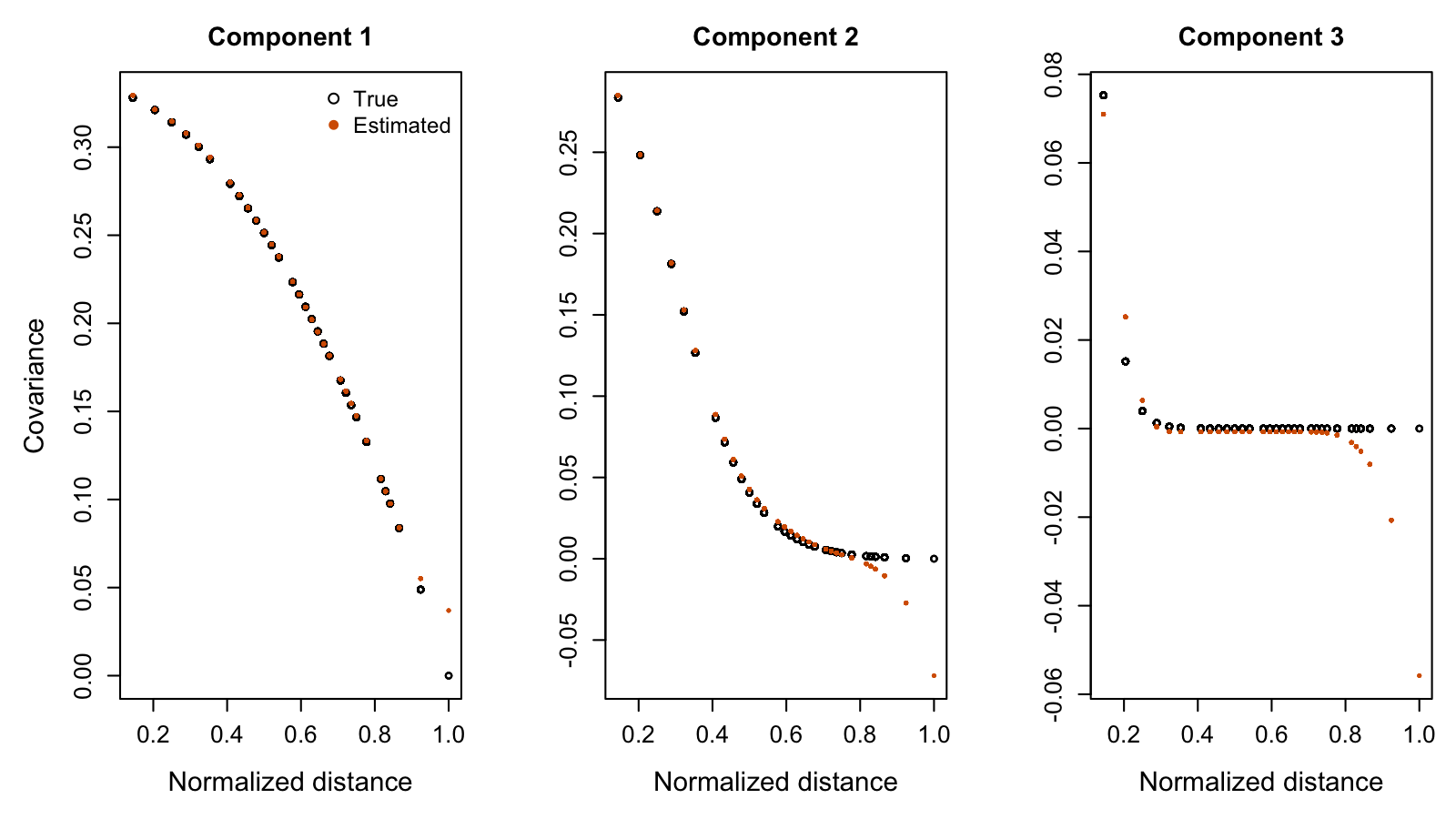}
    \caption{True and estimated off-diagonal spatial covariances for the three components as functions of normalized distance with scaling parameters $\hat{\alpha}_{1g}$ and $\hat{\alpha}_{2g}$. Open circles show the true values, and filled circles show the estimates.}
    \label{fig: Sim1 Cov vs Distance}
\end{figure}

 Among the 100 replicates, 99 achieved a perfect ARI of 1.0, while the remaining replicate yielded an ARI of 0.7388. Consequently, the mean ARI across all replicates was 0.9974, with a standard deviation of 0.0261. Figure~\ref{fig: Sim1 Cov vs Distance} compares the true and estimated spatial covariance functions for the three components in the displayed replicate.
The fitted spatial covariance closely follows the truth over most of the observed distance range, particularly for Components 1 and 2. The most pronounced discrepancies occur near the boundaries of the distance domain. Specifically, the fitted covariance for Component 2 turns slightly negative at the largest distances, while Component 3 exhibits deviations at both extremes. These edge effects stem from the non-uniform distribution of pairwise Euclidean distances on a finite grid. Since relatively few point pairs exist at these extreme distances, there is less empirical data available to accurately estimate the covariance function near the boundaries. Nevertheless, the rarity of these pairs ensures their impact on the overall recovery of the spatial covariance matrix remains minimal. Table~\ref{tab: MSE of cov} summarizes the entry-wise mean squared error (MSE) of the estimated full covariance matrices across all 100 replicates.

\begin{table}[ht]
\centering
\caption{Mean and standard deviation (SD) of the entry-wise MSE of the estimated full covariance matrices across 100 replicates.}
{\setlength{\tabcolsep}{6pt}
\renewcommand{\arraystretch}{1}
\begin{tabular}{c|ccc}
\hline
Group & 1 & 2 & 3 \\ 
\hline
Mean
& $8.2710\times10^{-5}$
& $4.5742\times10^{-4}$
& $3.3485\times10^{-3}$ \\ 
SD
& $7.0661\times10^{-5}$
& $3.3286\times10^{-3}$
& $3.2483\times10^{-2}$ \\ 
\hline
\end{tabular}
}
\label{tab: MSE of cov}
\end{table}

The comparatively large means and standard deviations for Groups 2 and 3 are driven primarily by the only replicate with an ARI below 1. Its MSEs of full covariance for these groups were 0.0334 and 0.3249, respectively. Excluding this replicate, the mean MSEs of the full covariance matrix for Groups 1--3 were $8.3011\times10^{-5}$, $1.2460\times10^{-4}$, and $1.0016\times10^{-4}$, respectively. Thus, covariance recovery was consistently accurate for the 99 replicates with perfect clustering, while the deterioration in covariance estimation was concentrated in the single replicate with imperfect clustering.

In summary, while the recovery of individual parameters and decay curves exhibits minor deviations due to boundary effects and scale ambiguity, the overall reconstruction of the full covariance matrices is highly accurate. These results demonstrate that the proposed spatial covariance structure offers significant flexibility to approximate different decay functions, including both quadratic and sigmoid profiles, within a unified framework.

\subsection{Simulation II - effect of spatial dimension}
To evaluate the performance of the proposed method as the spatial dimension increases, three regular two-dimensional grids of size $10\times10$, $14\times14$, and $20\times20$ were considered, corresponding to $p=100{,}196,$ and $400$ spatial locations, respectively. Each observation has a non-spatial dimension of $10$, resulting in vector dimensions of $1{,}000$, $1{,}960$, and $4{,}000$. In all scenarios, $N=1{,}000$ observations were generated from a three-component matrix normal mixture with mixing proportions $(0.5,0.3,0.2)$, zero component means, and three latent factors in the column covariance. The column covariance matrices were generated once and held constant across grid sizes, with factor loadings sampled from $\operatorname{Unif}(-1,1)$ and diagonal uniquenesses from $\operatorname{Unif}(0,1)$. For component $g$, the spatial row covariance was
\begin{equation*}
    \boldsymbol{\Xi}_g= \left[\exp\{-\lambda_g d_{ij}\}\right]_{i,j=1}^{p} + \mathbf{I}_p, \qquad (\lambda_1,\lambda_2,\lambda_3) = (3,10,30),
\end{equation*}
where $d_{ij}$ is the Euclidean distance between locations $i$ and $j$, normalized by the maximum distance on the grid. This construction gives a positive definite row covariance for all grid sizes and yields three distinct decay rates. For each dimension, 100 independent datasets were generated, and the MSFA was fitted using the true numbers of components and factors. The spatial decay function was modeled using cubic I-splines with six interior knots, with a common diagonal noise variance across spatial locations within each component.

All 100 fits converged at each grid size. As shown in Table~\ref{tab:sim2_dimension}, the mean ARI was 1.000 at $p=100$, 0.990 at $p=196$, and 0.963 at $p=400$. Perfect clustering was achieved in 100, 97, and 90 replicates, respectively. Thus, clustering accuracy remained high at a vector dimension of 4,000, although a slight decline was observed as the grid size increased. With the non-spatial dimension held constant, the fitted MSFA required 48 free covariance parameters per component, or 144 across the three components. In contrast, one unrestricted covariance matrix for a 4,000-dimensional vector would require 8,002,000 distinct free parameters.

Computational performance was evaluated by measuring individual fitting times in a separate sequential benchmark. For each grid size, five additional independent datasets were generated, and the MSFA was fitted once to each using a single worker on an Apple M4 computer with 16 GB RAM, resulting in 15 total fits. Elapsed fitting times did not include data generation. All 15 fits converged. The mean fitting times were 77.9, 675.4, and 1,632.9 seconds at $p=100$, 196, and 400, respectively. The corresponding mean times per EM iteration were 0.60, 2.05, and 7.48 seconds. The mean iteration counts were 130.0, 329.8, and 218.2.

\begin{table}[htbp]
\centering
\caption{Clustering accuracy across spatial dimensions. Results are based on 100 independent replicates per dimension.}
\label{tab:sim2_dimension}
\begin{tabular}{ccc}
\hline
Spatial locations $p$ & Vector dimension $pq$ & ARI (mean $\pm$ SD) \\
\hline
100 & 1,000 & $1.000 \pm 0.000$ \\
196 & 1,960 & $0.990 \pm 0.060$ \\
400 & 4,000 & $0.963 \pm 0.112$ \\
\hline
\end{tabular}
\end{table}

\subsection{Simulation III - selection of components and factor rank} \label{sec:sim4}
We considered a model selection setting on a $10\times10$ grid ($p=100$), with $q=10$, $N=1{,}000$, and a generating factor number of $r=4$. In the setting, data were generated from four equally weighted components. The spatial row covariances followed quadratic, two sigmoid, and logarithmic decay profiles, each with a common diagonal noise term. Unlike Simulation I, the four component mean matrices were distinct: one was zero, two were constant at $20$ and $-20$, and the fourth had entries drawn once from $\operatorname{Unif}(-20,20)$ and held fixed across replicates. The covariance parameters were also held fixed across replicates. For each of 100 independently generated datasets, we fitted all nine combinations of $G\in\{3,4,5\}$ components and $r\in\{3,4,5\}$ factors, using cubic I-splines with six knots and selecting the candidate with the largest BIC.

All 900 candidate fits converged. BIC selected the correct number of components in all 100 replicates and the joint generating configuration $(G,r)=(4,4)$ in 95. In the other five, it selected $(4,5)$, as shown in Table~\ref{tab:sim4-selection}.

\begin{table}[ht]
\centering
\caption{Frequencies of models selected by BIC across 100 independent replicates in the four-component setting. The generating model was $(G,r)=(4,4)$.}
\label{tab:sim4-selection}
\begin{tabular}{lccc}
\toprule
    & $G=3$ & $G=4$ & $G=5$\\
\midrule
        $r=3$ & 0 & 0 & 0 \\
        $r=4$ & 0 & 95 & 0 \\
        $r=5$ & 0 & 5 & 0 \\
\bottomrule
\end{tabular}
\end{table}

% \begin{table}[ht]
% \centering
% \renewcommand{\arraystretch}{1.8} % Increases row height to accommodate fractions
% \caption{Parameter Configuration for $\boldsymbol{\Xi}_g$ Matrices}
% \label{tab:xi_params}
% \begin{tabular}{@{}lll@{}}
% \toprule
% \textbf{Group} & \textbf{Decay Function} $h(x)$ & \textbf{Structure} \\ 
% \midrule
% $\boldsymbol{\Xi}_1$ & $h(x) = x^2$ & $\mathbf{J} - \mathbf{D} + \mathbf{I}$ \\
% $\boldsymbol{\Xi}_2$ & $h(x) = \frac{\sigma(10x-3) - \sigma(-3)}{\sigma(k-a) - \sigma(-a)}$ & $3\mathbf{J} - 2\mathbf{D} + \mathbf{I}$ \\
% $\boldsymbol{\Xi}_3$ & $h(x) = \frac{\sigma(30x-3) - \sigma(-3)}{\sigma(k-a) - \sigma(-a)}$ & $2\mathbf{J} - \mathbf{D} + \mathbf{I}$ \\
% $\boldsymbol{\Xi}_4$ & $h(x) = \ln(1 + (e-1)x)$ & $3\mathbf{J} - 2\mathbf{D} + \mathbf{I}$ \\ 
% \bottomrule
% \end{tabular}

% \vspace{0.5em}
% \footnotesize{
% \textit{Note:} $\sigma(z) = (1 + e^{-z})^{-1}$ is the standard sigmoid function. $J$ denotes a matrix of ones, $M$ is the matrix of transformed distances where $M_{ij} = f(d_{ij})$, and $I$ is the identity matrix.
% }
% \end{table}

\subsection{Simulation IV - comparison with alternative mixtures}
This simulation study evaluates the performance of the proposed MSFA model against three established clustering frameworks: MCLUST \citep{scrucca2016mclust}, PGMM \citep{mcnicholas08}, and MatrixMixtures \citep{tomarchio2022mixtures}. With a dimension of $5 \times 5 \times 5 \times 10$, data were generated under the MSFA framework assuming a two-component mixture with equal mixing proportions ($\pi_1 = \pi_2 = 0.5$). Within each component, the data generating model is driven by $r=3$ latent factors, utilizing sigmoid and logarithmic decay functions to model spatial dependence. The location tensor is fixed at zero for both groups. The scaling parameters $(\alpha_{1g}, \alpha_{2g})$ were specified as $(2,1)$ for the first cluster and $(3,2)$ for the second. The diagonal spatial variances $\boldsymbol{\gamma}_g$ were sampled randomly from the interval $[1,3]$. Due to the execution time of PGMM, we conducted 30 rather than 100 replicates to keep the total runtime of this comparison manageable. For all competing models, the number of mixture components was fixed at the true value ($G=2$). However, for the PGMM, the specific covariance constraint structure and the number of latent factors were not fixed. Instead, they were selected by optimizing the BIC over a factor range of $\{1, 2, \dots, 16\}$. For the MSFA, the candidate numbers of factors are $\{1,2,3,4,5\}$. The comparative performance and goodness-of-fit were assessed using the BIC.

The high dimensionality of the spatial covariance structure presented significant challenges for the competing methods. The matrix variate normal mixture model (MatrixMixtures) failed to fit the data. Despite the parsimony typically afforded by the Kronecker product decomposition, the computational burden of estimating full unstructured across-row and across-column covariance matrices proved prohibitive for a spatial system of this magnitude. MCLUST, while computationally stable, failed to capture the underlying group structure entirely. It selected a restrictive diagonal `EEI' model, resulting in an ARI of approximately $0$ and a BIC of $-5{,}782{,}774$, indicating a complete inability to model the complex spatial dependencies.

In contrast, PGMM achieved perfect clustering accuracy (ARI $= 1.0$), identifying a `CUU' model with $r=13$ latent factors. However, this accuracy came at the cost of substantial over-parameterization. The resulting BIC was $-5{,}362{,}820$, which is substantially poorer than the BIC of $-4{,}303{,}511$ achieved by the proposed MSFA. As summarized in Table~\ref{tab: compare}, these results demonstrate that for high-dimensional spatial data, the proposed MSFA framework achieves superior model fit with significantly greater parsimony than established alternatives.

\begin{table}[ht]
\centering
\caption{Comparison of clustering performance and model fit ($N=1000$, $G=2$) over 30 replicates.}
\label{tab: compare}
\resizebox{\textwidth}{!}{%
\begin{tabular}{lcccc}
\hline
Model & Specification / Structure & Status & ARI (mean) & BIC (mean) \\ \hline
MSFA & Spatial Factors ($r=3$) & Converged & 1.00 & $-4{,}303{,}511$ \\
PGMM & CUU ($r=13$) & Converged & 1.00 & $-5{,}362{,}820$ \\
MCLUST & EEI (Diagonal) & Converged & $\approx 0.00$ & $-5{,}782{,}774$ \\
MatrixMixtures & MVN & Failed & N/A & N/A \\ \hline
\end{tabular}%
}
\vspace{0.2cm}
\begin{minipage}{0.9\textwidth}
\footnotesize
\end{minipage}
\end{table}

\section{Real Data Analysis}
To demonstrate the practical utility of the MSFA, we apply the model to three real-world datasets involving high-dimensional spatial observations. 
% The first analysis uses Raman spectroscopy (RS) data from radiochromic films to investigate the impact of radiation dose on the underlying spatial covariance structures. The second application evaluates the model classification performance using the SpecTex hyperspectral texture database, aiming to differentiate textile samples based on their complex spatial-spectral properties.

\subsection{Raman spectroscopy analysis}
\begin{figure}[ht]
    \centering
    \includegraphics[width=1\linewidth]{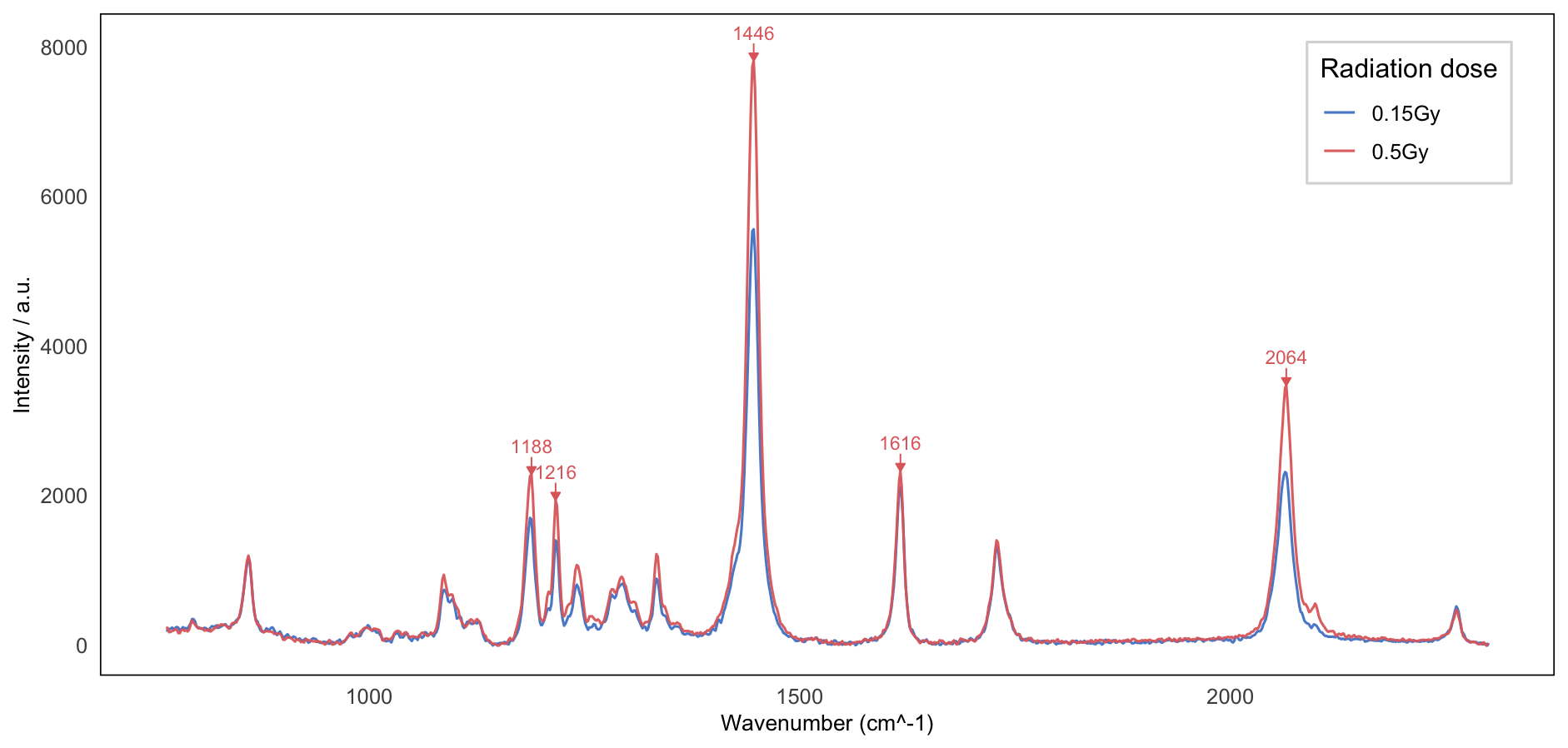}
    \caption{Raman spectra of dosimetric film at two different dose levels.}
    \label{fig: raman spec}
\end{figure}

To validate the practical effectiveness of the MSFA, we assess its performance using Raman spectroscopy (RS) data. RS measures relative molecular abundances and vibrational dynamics within an integrated sample. When coupled with a microscopic platform, RS provides spatial resolution on the order of micrometers \citep{mulvaney2000raman, das2011raman, orlando2021comprehensive}. The spectra analyzed in this study were obtained from measurements on EBT-3 radiochromic dosimeter films \citep{mcnairn2025exploring} used in radiation dosimetry. Upon exposure to ionizing radiation, the active layer between two matte polyester substrates undergoes polymerization, resulting in a color change proportional to the absorbed dose \citep{borca2013dosimetric}. RS provides a non-invasive, high-resolution approach for detecting this polymerization by monitoring peaks associated with the radiosensitive active elements, as illustrated in Figure~\ref{fig: raman spec}.

\begin{figure}
    \centering
    \includegraphics[width=1\linewidth]{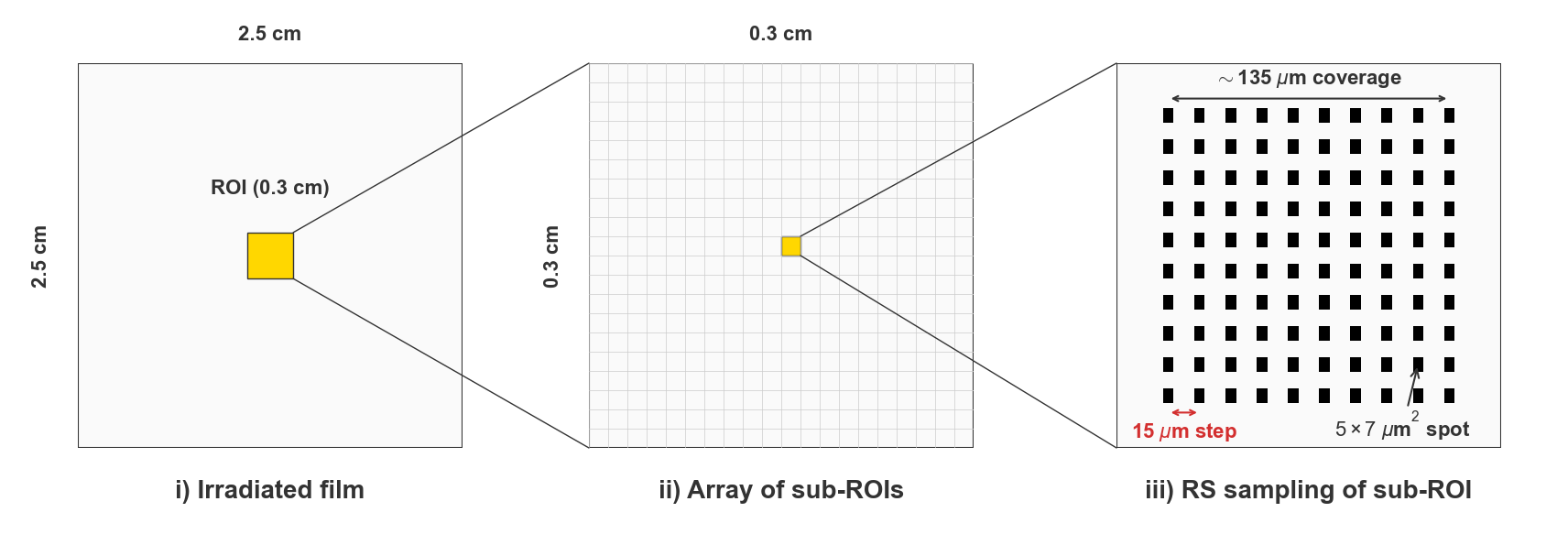}
    \caption{Schematic representation of the RS sampling protocol. (i) Macroscopic view of the irradiated EBT-3 film ($2.5 \times 2.5$ cm). (ii) The ROI comprises an array of 400 separate sub-ROIs (grids) arranged in a $20 \times 20$ pattern. (iii) Microscopic view of a sub-ROI, illustrating the $10 \times 10$ point-scan geometry.}
    \label{fig: Figure_1}
\end{figure}

\begin{figure}[!ht]
    \centering
    \includegraphics[width=0.75\linewidth]{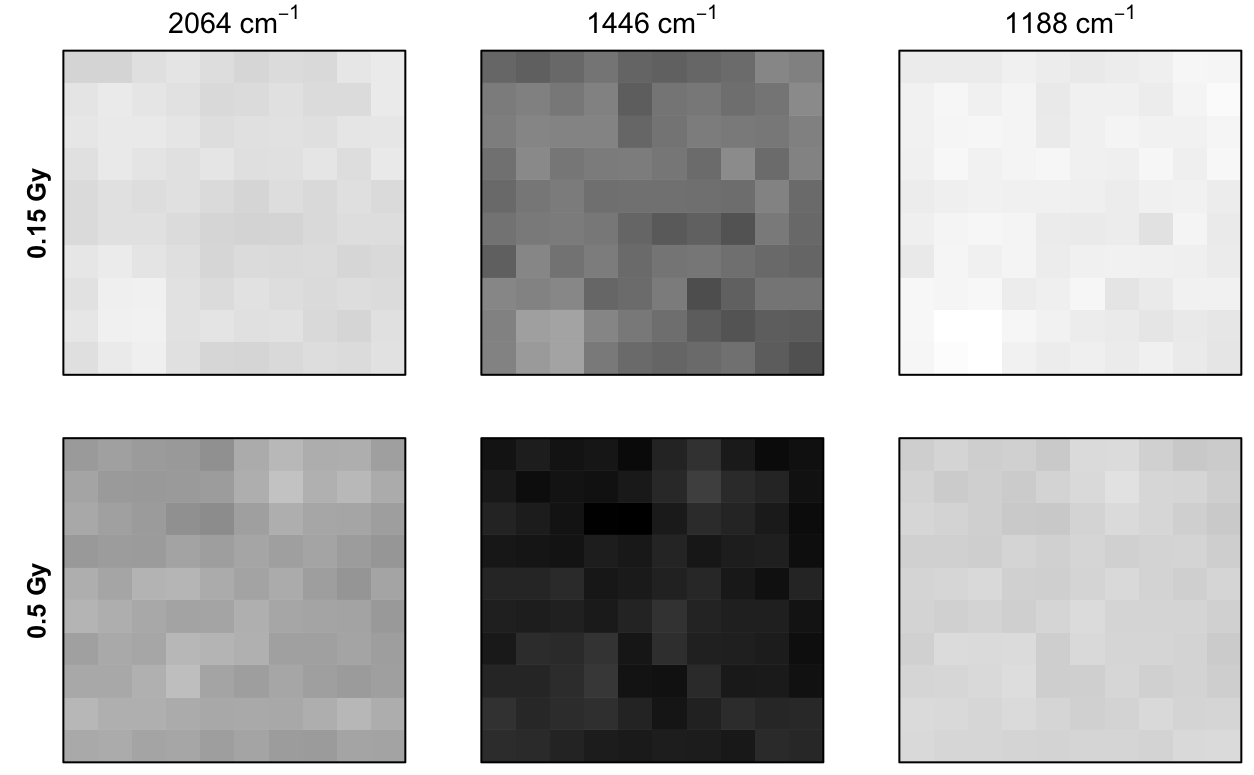}
    \caption{Visualization of three Raman peaks at $2064$ $\text{cm}^{-1}$, $1446$ $\text{cm}^{-1}$, and $1188$ $\text{cm}^{-1}$ measured on a $10 \times 10$ grid across two dosimetric films exposed to different radiation levels (0.5 Gy bottom, 0.15 Gy top).}
    \label{fig: examples of tensors two dose levels}
\end{figure}

Data collection followed a protocol where measurements were taken from a region of interest (ROI) partitioned into an array of distinct subsections of ROI (sub-ROIs), as shown in Figure~\ref{fig: Figure_1}. 
We acquired $40{,}000$ spectra across $400$ individual sampling grids for each film, resulting in a total of $80{,}000$ spectra across $800$ grids. Each sub-ROI utilized a $10 \times 10$ spatial array. The dataset includes observations from two films irradiated at doses of $0.15$ Gy and $0.50$ Gy, respectively. Spectra were recorded over a wavenumber range of $764$ to $2300$ $\text{cm}^{-1}$. For the analysis, we specifically targeted three peaks known to exhibit dose dependency, located at $2064$ $\text{cm}^{-1}$, $1446$ $\text{cm}^{-1}$, and $1188$ $\text{cm}^{-1}$ (see Figure~\ref{fig: raman spec}). Representative tensor observations corresponding to both dose levels are visualized in Figure~\ref{fig: examples of tensors two dose levels}.

As illustrated in Figure~\ref{fig: examples of tensors two dose levels}, the two dose groups are visually distinct, making clustering based solely on mean intensity values relatively straightforward. However, a more critical objective is to determine whether the groups exhibit distinct spatial dependency patterns. To investigate this, we group-centered the tensor observations, thereby removing the mean signal to isolate the underlying spatial covariance structure. Subsequently, we evaluated four variations of the MSFA model with $G=2$ and $r=1$ to test specific hypotheses regarding the spatial and noise structures. These variations involved applying constraints of either homoscedastic diagonal spatial noise ($\boldsymbol{\gamma}_g = \alpha_{3g}\mathbf{1}$), common spatial covariance matrices across groups ($\boldsymbol{\Xi}_1 = \boldsymbol{\Xi}_2$), or both simultaneously, with the degree of I-splines $m=3$ and number of knots $k=6$.

\begin{figure}[!ht]
    \centering
    \includegraphics[width=1\linewidth]{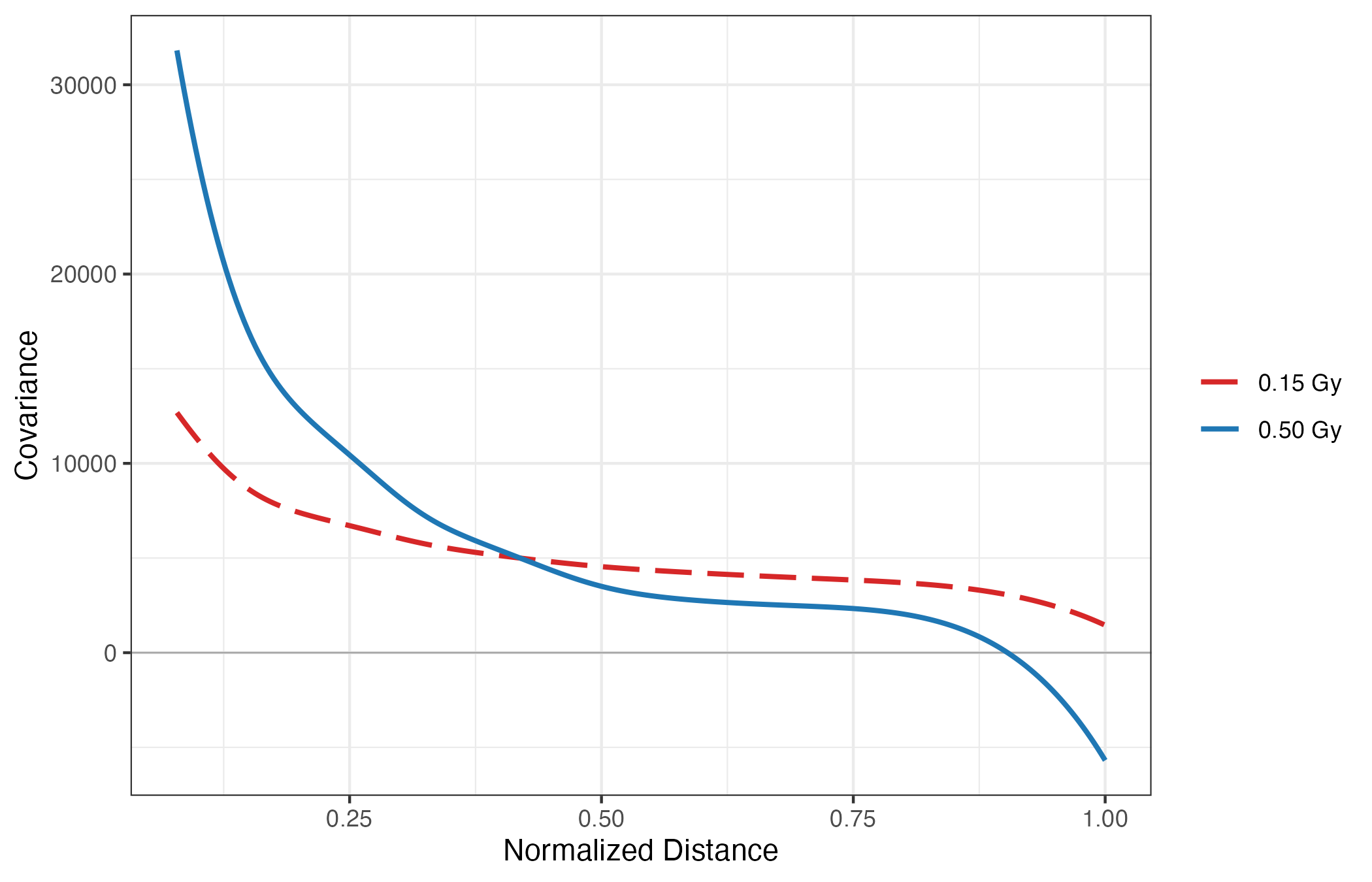}
    \caption{Estimated spatial covariance versus normalized distance for the two radiation dosage groups. The red dashed line represents the low-dose (0.15 Gy) group, while the blue solid line represents the high-dose (0.50 Gy) group.}
    \label{fig: Real fitted curve2}
\end{figure}

Based on the BIC comparison, the model under the constraint of homoscedastic diagonal spatial noise ($\boldsymbol{\gamma}_g = \alpha_{3g}\mathbf{1}$) is preferred. This configuration yielded a BIC of $-3{,}025{,}352$ and achieved a high clustering accuracy with an ARI of $0.955$. The two films exhibit different fitted spatial covariance patterns, as shown in Figure~\ref{fig: Real fitted curve2}, although this study cannot uniquely attribute the difference to dose. Furthermore, an analysis of the estimated full covariance matrices reveals a disparity in total variability between the groups. The high-dose group exhibits considerably greater variability, with the trace of its full covariance matrix, $\operatorname{tr}(\hat{\boldsymbol{\Omega}}_2 \otimes \hat{\boldsymbol{\Xi}}_2) = 48{,}277{,}521$, exceeding that of the low-dose group, $\operatorname{tr}(\hat{\boldsymbol{\Omega}}_1 \otimes \hat{\boldsymbol{\Xi}}_1) = 29{,}897{,}539$.
For comparison, PGMM, MCLUST, and MatrixMixtures were applied to the centered data. Among these alternatives, only the PGMM converged to a valid solution. However, its clustering performance and goodness-of-fit were substantially poorer than those of the proposed model. With the number of latent factors ranging from 1 to 12, the PGMM yielded an ARI of $0.74$ and a BIC of $-3{,}290{,}751$, which is substantially lower than the BIC achieved by the MSFA. Furthermore, the PGMM failed to identify a parsimonious structure, since its BIC score exhibited a strictly increasing trend with the addition of latent factors, suggesting that the search range was insufficient to identify an interior optimum.

%However, only PGMM yielded meaningful convergence. Notably, the PGMM produced a lower BIC of $-3,290,751$ compared to the MSFA model. Additionally, the BIC for PGMM strictly increased with the number of factors, resulting in a failure to identify an optimal number of factors.

\subsection{SpecTex hyperspectral texture analysis}
\begin{figure}[ht]
    \centering
    \includegraphics[width=0.8\linewidth]{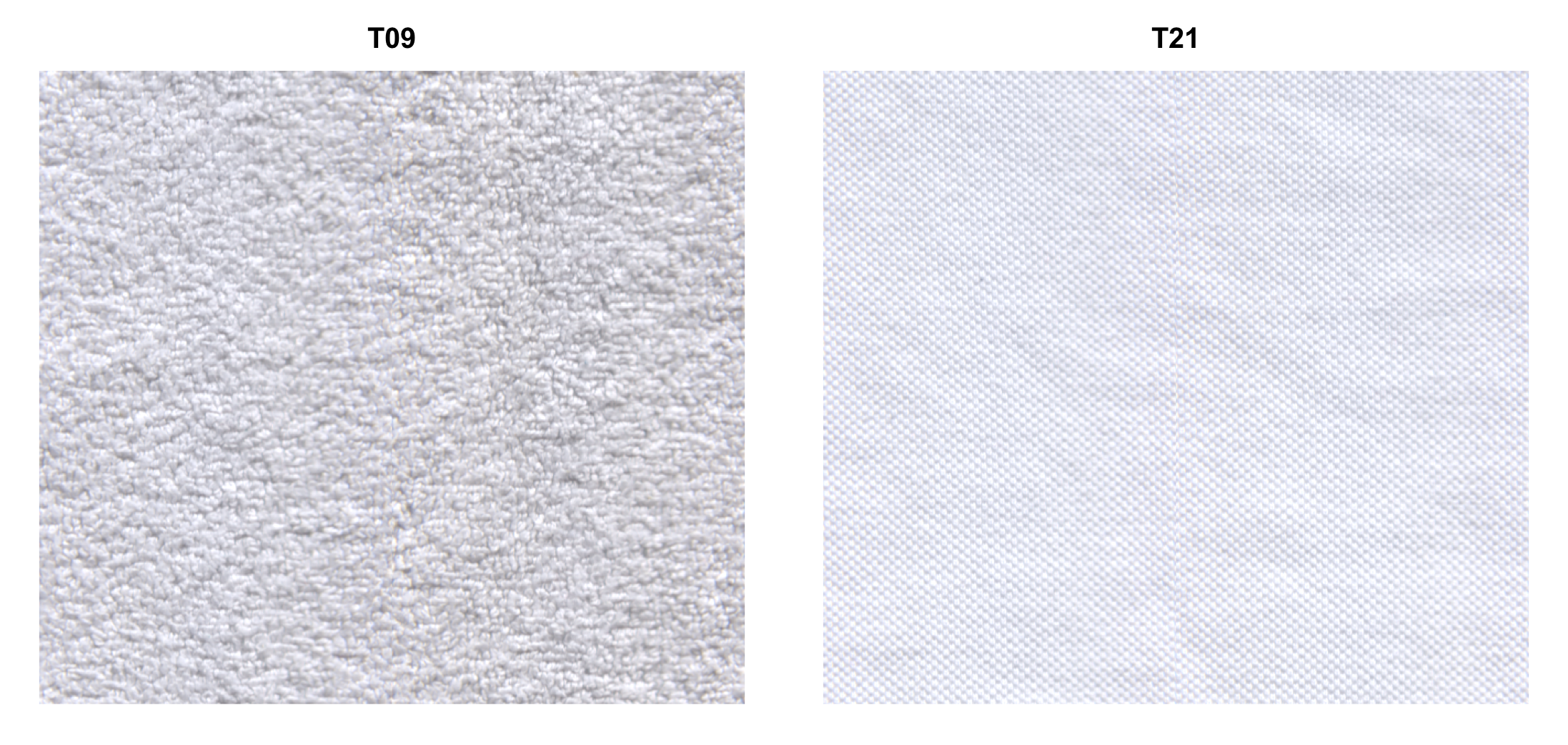} % Adjusted width
    \caption{Visual representation of the two textile samples selected from the SpecTex database: `T09' (left) and `T21' (right).}
    \label{fig: Real2-Viz examples}
\end{figure}

The second real-data evaluation employs the SpecTex Hyperspectral Texture Image Database \citep{mirhashemi2018introducing}, a benchmark collection developed by the University of Eastern Finland for spectral texture analysis. This dataset comprises hyperspectral reflectance images of 60 distinct textile samples, capturing a diverse range of weave patterns and material compositions that require both spatial and spectral discrimination. The standardized $10$ nm dataset provides spectral data across the visible range ($400$--$780$ nm) at $10$ nm intervals, yielding a total of $39$ spectral bands.
% In the standardized 10 nm dataset, spectral data is across the visible range ($400$--$780$ nm) at $10$ nm intervals, resulting in $39$ spectral bands. 
Each observation is structured as a high-dimensional tensor with a spatial resolution of $640 \times 640$ pixels. For this analysis, we selected two distinct textile samples, labeled `T09' and `T21', as illustrated in Figure~\ref{fig: Real2-Viz examples}.

Prior to modeling, we implemented a preprocessing step to address the vertical striping artifacts, which are common in push-broom scanner data. This procedure normalizes each spatial column to match the global properties of the spectral band, effectively correcting non-uniform sensor responses while preserving the underlying textural information. Subsequently, to reduce spectral redundancy while retaining multivariate complexity, we subsampled the data to five spectral bands (indices 1, 10, 20, 30, and 39) uniformly distributed across the visible range. Following artifact correction and spectral selection, the images were tessellated into non-overlapping spatial patches to generate a sufficient sample size. Each $640 \times 640$ image is partitioned into grids of size $32 \times 32$ pixels, yielding 400 observations per texture class and a total dataset size of $N=800$. After vectorization, the data were organized into a set of $N=800$ matrix variate observations, each with dimensions $1{,}024 \times 5$ (corresponding to $32 \times 32$ spatial locations and $5$ spectral bands).

\begin{figure}[ht]
    \centering
    \includegraphics[width=0.8\linewidth]{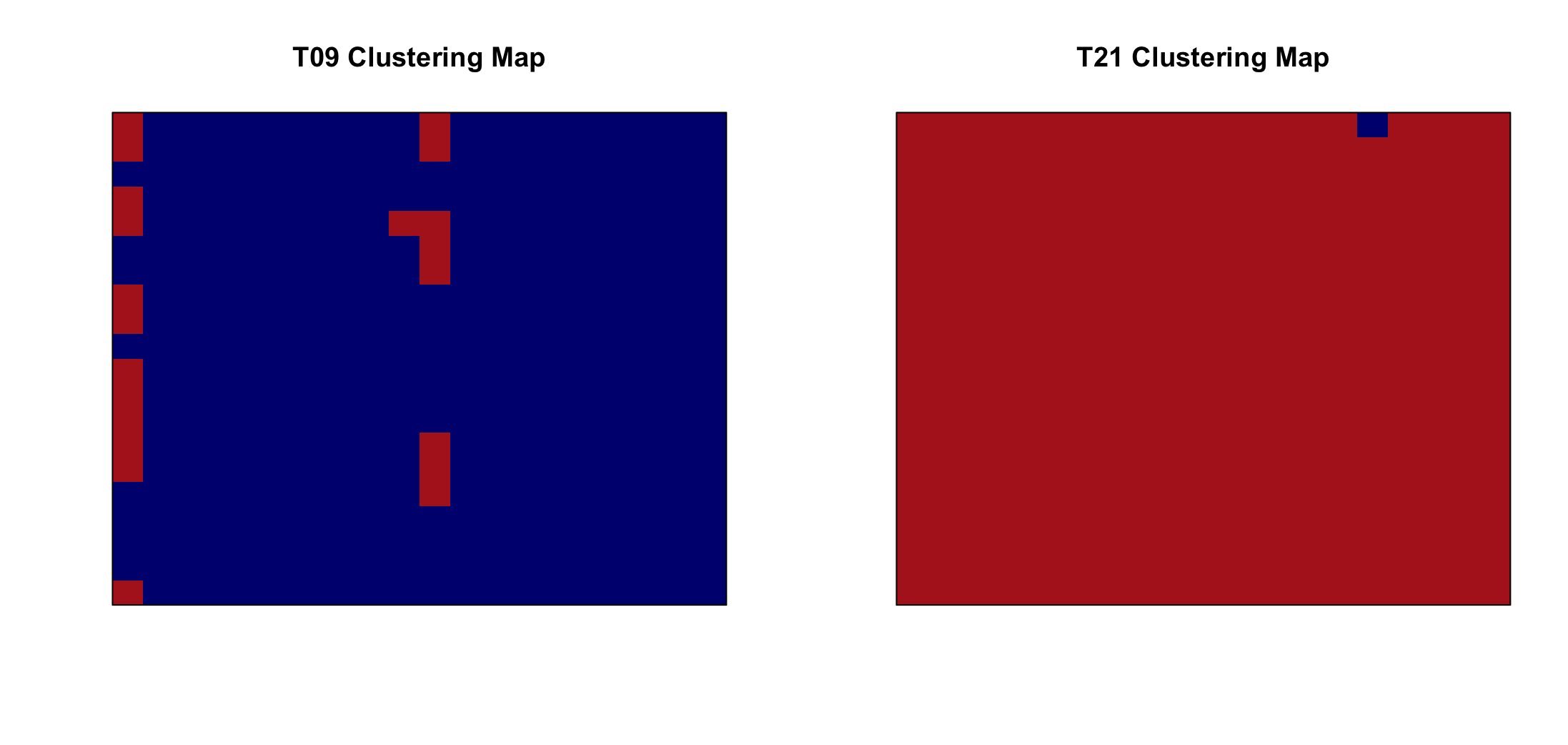}
    \caption{Clustering results for the SpecTex dataset. The maps visualize the cluster assignments for the patches from sample `T09' (left) and `T21' (right).}
    \label{fig: Real2 Clustering}
\end{figure}

We applied the proposed MSFA to the combined dataset to evaluate its capacity to differentiate the texture patterns in an unsupervised manner. We fit models with the number of factors $r \in \{1, 2\}$ and constraints including $\boldsymbol{\gamma}_g = \alpha_{3g}\mathbf{1}$ and $\boldsymbol{\Xi}_1 = \boldsymbol{\Xi}_2$. All models utilized cubic I-splines with 10 knots. Based on the BIC, the optimal model specification was identified as having two latent factors ($r=2$) with group-specific spatial covariance matrices ($\boldsymbol{\Xi}_1 \neq \boldsymbol{\Xi}_2$) and homoscedastic diagonal spatial noise ($\boldsymbol{\gamma}_g = \alpha_{3g}\mathbf{1}$). This suggests that the two textiles exhibit distinct underlying spatial structures that cannot be captured accurately by a common covariance model. The model achieved an ARI of $0.8929$. The resulting classification maps, shown in Figure~\ref{fig: Real2 Clustering}, demonstrate that the MSFA effectively separates the two texture classes with few misclassifications.

\begin{figure}[ht]
    \centering
    \includegraphics[width=0.8\linewidth]{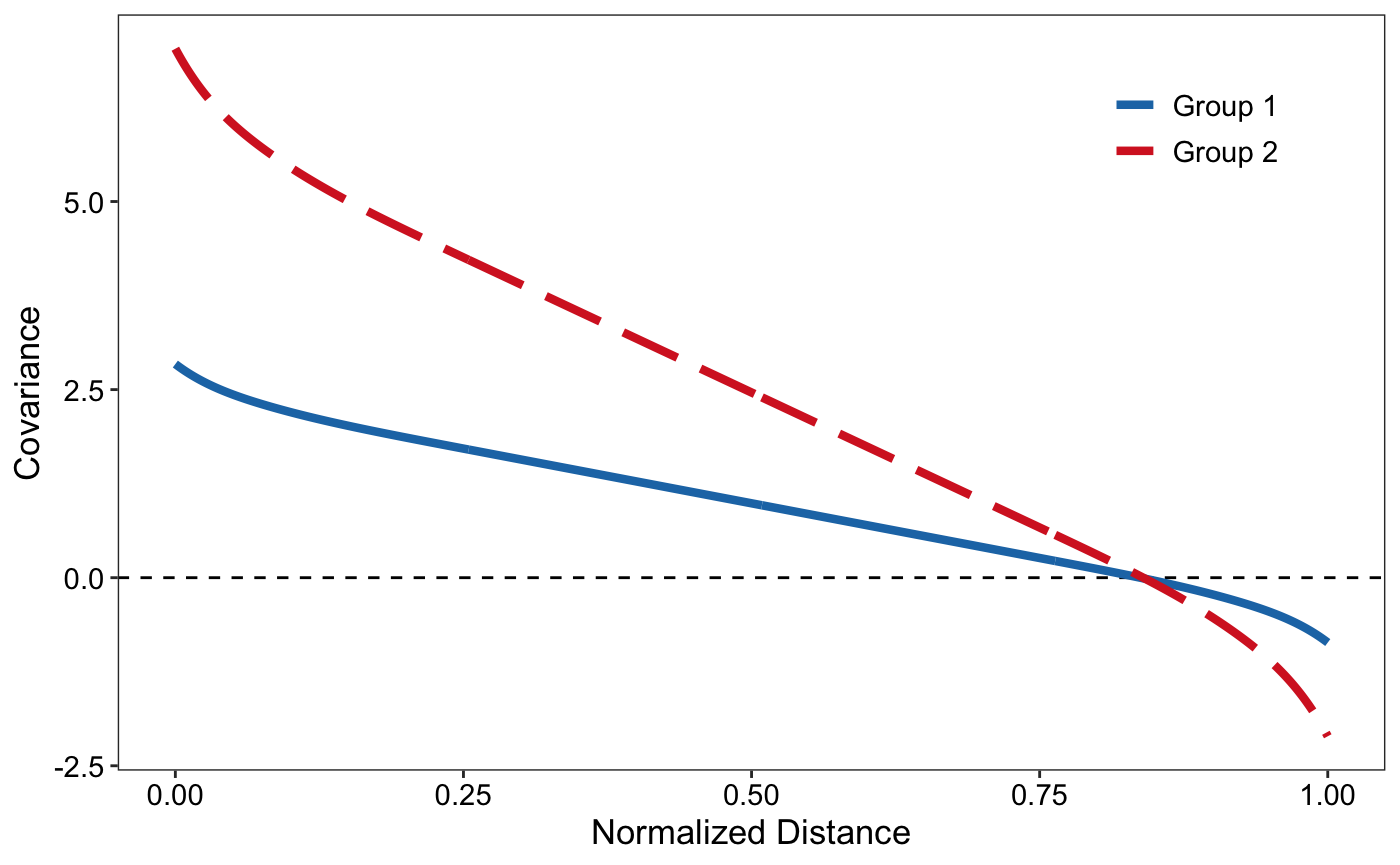}
    \caption{Estimated spatial covariance functions for the identified clusters. The curves illustrate the decay of spatial dependence over distance for Group 1 (blue solid) and Group 2 (red dashed).}
    \label{fig: Real2 Spatial}
\end{figure}

The estimated spatial covariance curves are visualized in Figure~\ref{fig: Real2 Spatial}. The fitted spatial variability differs between the two groups, as illustrated by the estimated scaling parameters $\hat{\alpha}_{21}=2.75427$ for Group 1 and $\hat{\alpha}_{22}=6.8087$ for Group 2. 

\subsection{Salinas hyperspectral dataset analysis}

\begin{figure}[ht]
    \centering
    \includegraphics[width=0.5\linewidth]{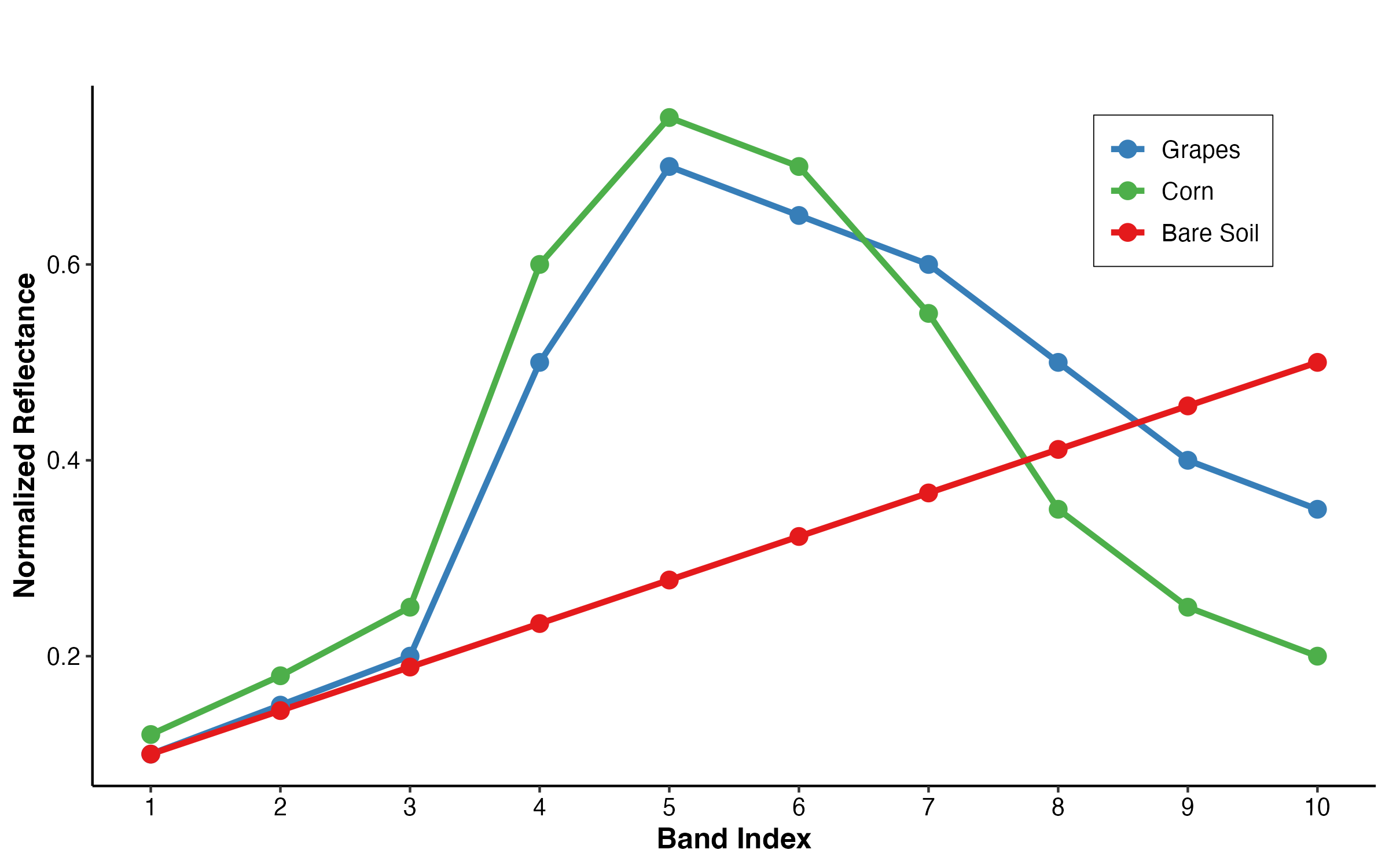}
    \caption{Mean spectral profiles of the three classes (Grapes, Corn, and Bare Soil).}
    \label{fig: Figure1_Spectra}
\end{figure}

\begin{figure}[ht]
    \centering
    \includegraphics[width=1\linewidth]{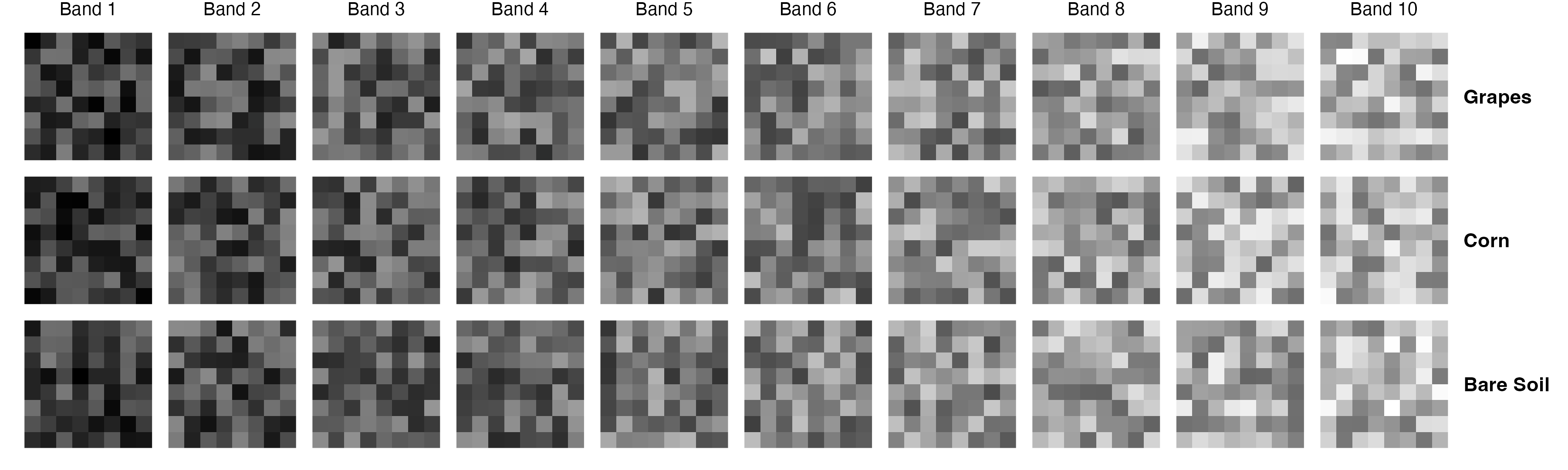}
    \caption{Visualization of representative matrix variate observations of the three classes (Grapes, Corn, and Bare Soil).}
    \label{fig: Figure2_Grid_White}
\end{figure}

To further assess the performance of the MSFA on real-world data, we utilized the Salinas Valley hyperspectral dataset \citep{salinas_dataset}, acquired by the AVIRIS sensor over agricultural regions in California. The full scene covers an area of $512 \times 217$ pixels with a spatial resolution of 3.7 meters and originally comprises 224 spectral bands. The mean spectral profiles of the three selected classes are illustrated in Figure~\ref{fig: Figure1_Spectra}.
For this analysis, we focused on distinguishing three specific classes, namely, grapes, corn, and bare soil. The data were normalized to the $[0,1]$ range, then uniformly subsampled to 10 bands within the reliable spectral range to reduce dimensionality. To capture local spatial correlation structures, we extracted non-overlapping patches of $8 \times 8$ pixels, strictly applying a purity constraint so that each patch contained pixels from only one class, as shown in Figure~\ref{fig: Figure2_Grid_White}. These patches were then vectorized, resulting in a dataset of $171$ matrix variate observations with dimensions $64 \times 10$.
We applied the proposed MSFA to the preprocessed dataset, evaluating models with $r \in \{1, \dots, 4\}$ using I-splines of degree 3 with 10 knots. Models with and without constraint $\boldsymbol{\gamma}_g = \alpha_{3g}\mathbf{1}$ were fitted. For benchmarking, we also fitted MCLUST on the flattened feature vectors. The proposed MSFA with $r=3$ and $\boldsymbol{\gamma}_g = \alpha_{3g}\mathbf{1}$ was selected by the BIC, yielding an ARI of $0.8918$ and a BIC of $1{,}163{,}807$. In comparison, MCLUST with a `VVI' covariance structure achieved an ARI of $0.8657$ with a lower BIC of $802{,}685$. While MCLUST achieves a good clustering result, the BIC of the MSFA indicates a much better model fit. Furthermore, with a spatial grid of $64$ locations and $10$ spectral bands, a location matrix involves $640$ free parameters per group, which is much larger than the number of observations $171$. To mitigate this, we imposed a spatially invariant mean constraint $\mathbf{M}_g = \mathbf{1}\boldsymbol{\mu}_g^\prime$, where $\boldsymbol{\mu}_g$ is a $q$-dimensional mean vector, assuming that the mean spectral signature is constant across the spatial patch. Under this constraint, the number of mean parameters is reduced to $q=10$ per group. The BIC still favored the specification with $r=3$ and $\boldsymbol{\gamma}_g = \alpha_{3g}\mathbf{1}$. In this dataset, the mean constraint increased ARI from 0.8918 to 0.9790 and BIC from 1,163,807 to 1,172,742. Only two observations were misclassified.

\begin{figure}[ht]
    \centering
    \includegraphics[width=1\linewidth]{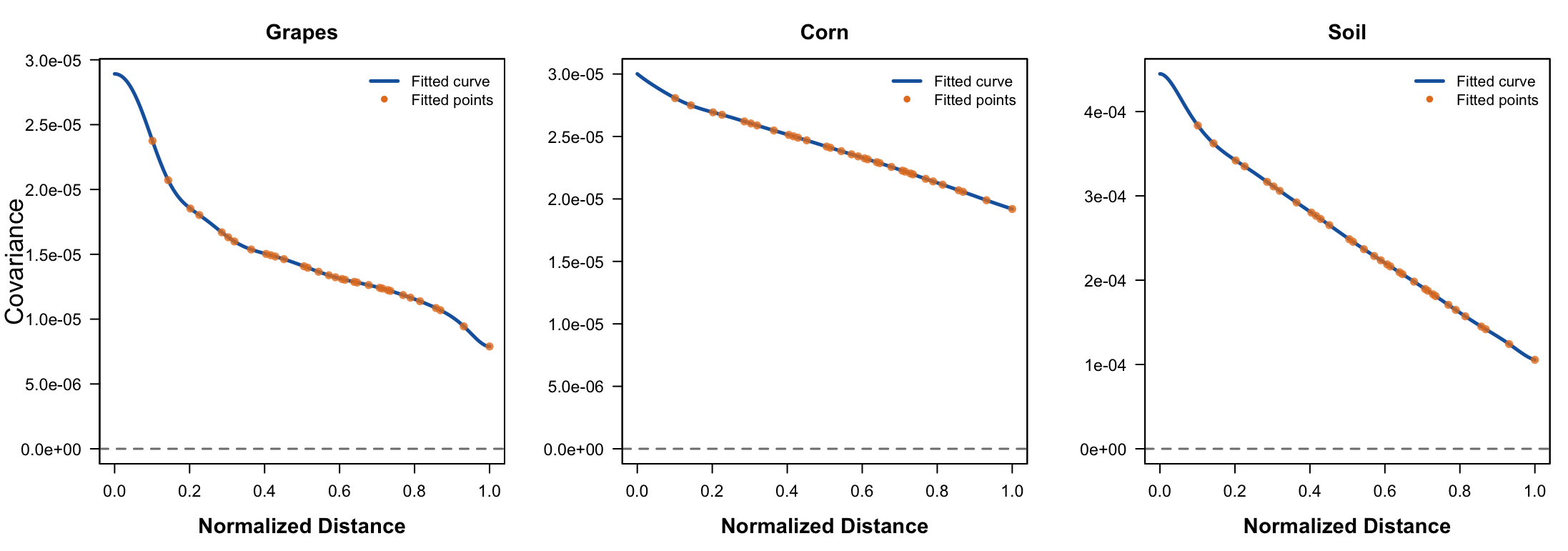}
    \caption{Estimated spatial covariance functions for the three classes (Grapes, Corn, Soil). The plots illustrate the covariance as a function of normalized distance. }
    \label{fig: Real3_curve}
\end{figure}

% In terms of the spatial pattern, Table~\ref{tab:alpha_transposed} summarizes the estimated linear spatial parameters $\hat{\boldsymbol{\alpha}}_g$.
% \begin{table}[htbp]
%   \centering
%   \caption{Estimated linear spatial covariance parameters $\hat{\boldsymbol{\alpha}}_g$.}
%   \label{tab:alpha_transposed}
%   \begin{tabular}{c|ccc}
%     \toprule
%     & $\alpha_{1g}$ & $\alpha_{2g}$ & $\alpha_{3g}$ \\
%     \midrule
%     Grapes & 0.6669  & 0.4891 & 0.8477 \\
%     Corn & 1.2316  & 0.9385 & 1.1932 \\
%     Soil & 0.0229 & 0.0167 & 0.0270 \\
%     \bottomrule
%   \end{tabular}
% \end{table}
In terms of the spatial pattern, the spatial covariance is visualized in Figure~\ref{fig: Real3_curve} as a function of normalized distance. The three classes exhibit fundamentally different spatial signatures, both in magnitude and shape. These distinct covariance curves provide evidence that the MSFA successfully disentangles the underlying spatial patterns of the observations, providing a mechanism for discrimination beyond simple spectral mean differences.

\section{Summary} \label{sec:summary}
We introduced a mixture of spatial factor analyzers for tensor-variate data. For fixed spline hyperparameters, the spatial covariance uses a number of parameters that grows linearly with the number of locations. In particular, it remains constant when all the locations share common diagonal noise. Factor analysis provides additional parsimony in the non-spatial dimension. Simulation studies and three applications illustrate the practical ability of the model to distinguish fitted spatial dependence patterns.

Analogous to the Parsimonious Gaussian Mixture Models (PGMM), the MSFA framework can be expanded into a broader family of models by imposing specific constraints—such as $\boldsymbol{\Lambda}_g = \boldsymbol{\Lambda}$, $\boldsymbol{\Psi}_g = \boldsymbol{\Psi}$, or $\boldsymbol{\Psi}_g = \psi_g \mathbf{I}$—on the non-spatial covariance components. These constraints would allow for a more parsimonious representation of the dependencies among the non-spatial variables. Moreover, since the spatial covariance structure in MSFA depends only on Euclidean distance, it may be misspecified on data with strong directional dependence, affecting clustering. However, for data on regular Cartesian grids, separate decay functions along the coordinate axes could be combined through a sum of Kronecker product terms. This would accommodate axis-aligned anisotropy beyond the current radial distance formulation.

\subsection*{Funding}
The authors acknowledge the support of the Government of Canada's New Frontiers in Research Fund (NFRF) [funding reference number NFRFE-2021-00312], Natural Sciences and Engineering Research Council (NSERC) of Canada (funding reference number RGPIN-2022-04897 (SM), RGPIN-2024-05355 (RMT), RGPIN-2020-07232 (AJ), RGPIN-2021-03812 (SS), RGPIN-2020-04646 (JLA)), the Canada Research Chairs program (RMT and SS).

\newpage
\bibliographystyle{apalike}
\bibliography{EntireLibrary}

\end{document}